\documentclass[aip,reprint]{revtex4-1}

\usepackage{graphicx}
\usepackage{blindtext}
\usepackage{siunitx}
\usepackage{multirow}
\usepackage{booktabs}
\usepackage{array}
\usepackage{amsmath}
\usepackage{textcomp}
\usepackage{url}
\usepackage[normalem]{ulem}

\begin{document}

\title{Temperature-dependent mechanical losses of Eu$^{3+}$:Y$_{2}$SiO$_{5}$ for spectral hole burning laser stabilization}

\author{Nico Wagner}
\email[]{nico.wagner@tu-braunschweig.de}
\affiliation{Institut für Halbleitertechnik, Technische Universität Braunschweig, Hans-Sommer-Str. 66, Braunschweig, 38106 Germany}
\affiliation{Laboratory for Emerging Nanometrology, Langer Kamp 6a-b, Braunschweig, 38106 Germany}
\author{Johannes Dickmann}
\affiliation{Institut für Halbleitertechnik, Technische Universität Braunschweig, Hans-Sommer-Str. 66, Braunschweig, 38106 Germany}
\affiliation{Laboratory for Emerging Nanometrology, Langer Kamp 6a-b, Braunschweig, 38106 Germany}
\author{Bess Fang}
\author{Michael T. Hartman}
\affiliation{LNE-SYRTE, Observatoire de Paris, Université PSL, CNRS, Sorbonne Université - Paris, France}
\author{Stefanie Kroker}
\affiliation{Institut für Halbleitertechnik, Technische Universität Braunschweig, Hans-Sommer-Str. 66, Braunschweig, 38106 Germany}
\affiliation{Laboratory for Emerging Nanometrology, Langer Kamp 6a-b, Braunschweig, 38106 Germany}
\affiliation{Physikalisch-Technische Bundesanstalt, Bundesallee 100, 38116 Braunschweig, Germany}

\begin{abstract}
We investigate the mechanical loss characteristics of Eu$^{3+}$:Y$_2$SiO$_5$—a promising candidate for ultra-low-noise frequency stabilization through the spectral hole burning technique.
Three different mechanical oscillators with varying surface-to-volume ratios and crystal orientations are evaluated.
In this context, we perform mechanical ringdown and spectral measurements spanning temperatures from room temperature down to \SI{15}{\K}.
By doing so, we measure a maximum mechanical quality factor of $Q=\num{3676}$, corresponding to a loss angle of $\phi=\num{2.72e-4}$.
For a spectral hole burning laser stabilization experiment at \SI{300}{\milli\K}, we can estimate the Allan deviation of the fractional frequency instability due to Brownian thermal noise to be below $\sigma_{\delta \nu/\nu_0} = \num{2.5e-18}$, a value lower than the estimated thermal-noise limit of any current cavity-referenced ultra-stable laser experiment.
\end{abstract}

\maketitle

\section{Introduction}
Lasers with ultra-high frequency stability are pivotal across various applications, including the advancement of optical lattice clocks \cite{schioppo2017ultrastable, tyumenev2016comparing, lisdat2016clock, ushijima2015cryogenic, nicholson2015systematic}, interferometric gravitational wave detectors \cite{abbott2016observation,Harry2006,Hild2011,Adhikari2020}, and the generation of ultra-low phase noise microwave signals.~\cite{xie2017photonic}
State-of-the-art ultra-stable lasers presently employ transmission peaks of high-finesse Fabry-Perot cavities as their frequency references. 
Brownian thermal noise is a significant inherent limitation to the stability in such systems, particularly if they are operated at room temperature, limiting their relative frequency stability to approximately $10^{-16}$.\cite{Numata2004}
To significantly enhance  performance, substantial modifications of the experiments are imperative. These include enhancing the cavity mode volume \cite{Haefner2015}, integrating crystalline materials for the spacer and mirror components \cite{cole2013tenfold, Kedar2023}, meta-mirrors \cite{dickmann2023experimental, Kroker2017}, employing cryogenic temperatures\cite{kessler2012sub}, or combining these strategies.

An alternative approach for frequency stabilization involves laser locking to a narrow optical transition in a rare earth ion-doped crystalline matrix.
This is achieved by spectral hole burning (SHB), i.e., locking the laser to spectral features created through photo-induced processes in these crystals.
Impressively narrow spectral features, as fine as $\SI{1}{\kHz}$, have been observed in Eu$^{3+}$:Y$_2$SiO$_5$ (Eu:YSO) crystals.~\cite{thorpe2011frequency}
Photon-echo experiments have even suggested the potential for achieving structures as narrow as a few 100 Hz.~\cite{equall1994ultraslow}
Demonstrations using this method have exhibited short-term stability of $6 \times 10^{-16}$ between 2 and 10 seconds \cite{thorpe2011frequency}, thus holding the promise of superior performance compared to conventional Fabry-Perot techniques.

Despite the extensive research in this field, a critical gap exists in the understanding of the mechanical losses of Eu:YSO, which could potentially curtail the achievable frequency stability attributable to Brownian thermal noise.
Ohta et al.~\cite{Ohta2021} examined the optomechanical interaction involving erbium ions within bulk YSO crystals structured as a mechanical resonator.
By analyzing the frequency-dependent displacement, they determined a mechanical quality factor of $Q=2500$ at a temperature of \SI{4}{\K}. 

To explore application of Eu:YSO at other temperatures, this study presents the measurement of temperature-dependent mechanical losses of Eu:YSO, spanning a range from room temperature down to \SI{15}{\K}. 
Moreover, we study the implications for ultra-stable lasers that are frequency-locked via the technique of SHB with Eu:YSO crystals.

\section{Experiment}
We investigate three Eu:YSO samples with different surface-to-volume ratio and crystal orientation, as detailed in Tab.~\ref{tab:samples}.
The samples are made from a bulk Eu$^{3+}$:Y$_2$SiO$_5$ crystal grown from the melt by the Czochralski method, and cut along the crystallographic $b$ axis as well as the dielectric $D_1$ and $D_2$ axis into a \SI{10}{\milli\meter} $\times$ \SI{15}{\milli\meter} $\times$ \SI{3}{\milli\meter} cuboid, with the largest facets perpendicular to the crystallographic $b$ axis. Due to the hardness (Young's modulus of 135 GPa at room temperatures) and its chemical inertness, patterning microscopic structures via, for example, focused ion beam etching is possible but time-consuming.~\cite{Motte2019}

\begin{table*}[tb]
\caption{\label{tab:samples}Dimensions of the three samples A, B, and C. The length $L$ ($l$), width $w$, and thickness $H$ ($t$) are given for the clamping block (oscillator). The crystal orientation for Eu:YSO can be expressed as the $D_1$ or $D_2$ axis, which are perpendicular to each other~\cite{YSO_axis}, and is specified for the axis bending in the flexural mode.}
\begin{ruledtabular}
    \begin{tabular}{lccccccc}
        \multirow{2}{*}{} & \multicolumn{3}{c}{Clamping} & \multicolumn{3}{c}{Oscillator} & \multirow{2}{*}{Crystal orientation} \\
        \cline{2-4} \cline{5-7}
        & $L$ & $w$ & $H$ & $l$ & $w$ & $t$ & \\
        \hline
        Sample A & \SI{6}{\milli\meter} & \SI{6}{\milli\meter} & \SI{0.85}{\milli\meter} & \SI{6}{\milli\meter} & \SI{6}{\milli\meter} & \SI{0.35}{\milli\meter} & $D_2$ along neutral axis\\
        Sample B & \SI{8.68}{\milli\meter} & \SI{10}{\milli\meter} & \SI{4.39}{\milli\meter} & \SI{10}{\milli\meter} & \SI{10}{\milli\meter} & \SI{1.05}{\milli\meter} & $D_2$ along neutral axis\\
        Sample C & \SI{7.49}{\milli\meter} & \SI{10}{\milli\meter} & \SI{3.23}{\milli\meter} & \SI{9}{\milli\meter} & \SI{10}{\milli\meter} & \SI{0.80}{\milli\meter} & $D_1$ along neutral axis\\
        \end{tabular}
    \end{ruledtabular}
\end{table*}

In order to reduce surface roughness induced mechanical loss~\cite{Nawrodt2013}, we have employed mechanical polishing methods instead to produce a thin resonator plate. For sample A, which is relatively thin, a step of $\SI{500}{\micro\meter}$ height is first milled away over an area of about ${\SI{10}{\milli\meter} \times \SI{10}{\milli\meter}}$ rectangular surface area to create a thicker zone for clamping and a thinner zone as the resonator. The lap surface, then the unmilled surface opposite to it, are subsequently polished down to a roughness of a few nanometer RMS, and a surface height variation of the order of \SI{100}{\nano\meter} at low spatial frequencies across the entire surface measured using an interferometric microscope (VICO NT1100). Polishing the unmilled surface also allows to adjust the overall thickness of the oscillator. As a final step, frailed edges are diced away to ensure the regularity of the oscillator's shape. The final dimension of sample A used for the loss measurements is about \SI{6}{\milli\meter} $\times$ \SI{6}{\milli\meter} $\times$ \SI{350}{\micro\meter}. An attempt to make thinner oscillators resulted in a fragile structure which eventually broke before any experiments could be carried out. Samples B and C were machined in a similar way, adapting the dimensions accordingly. 

\begin{figure}[tb]
    \centering
    \includegraphics[width=\linewidth]{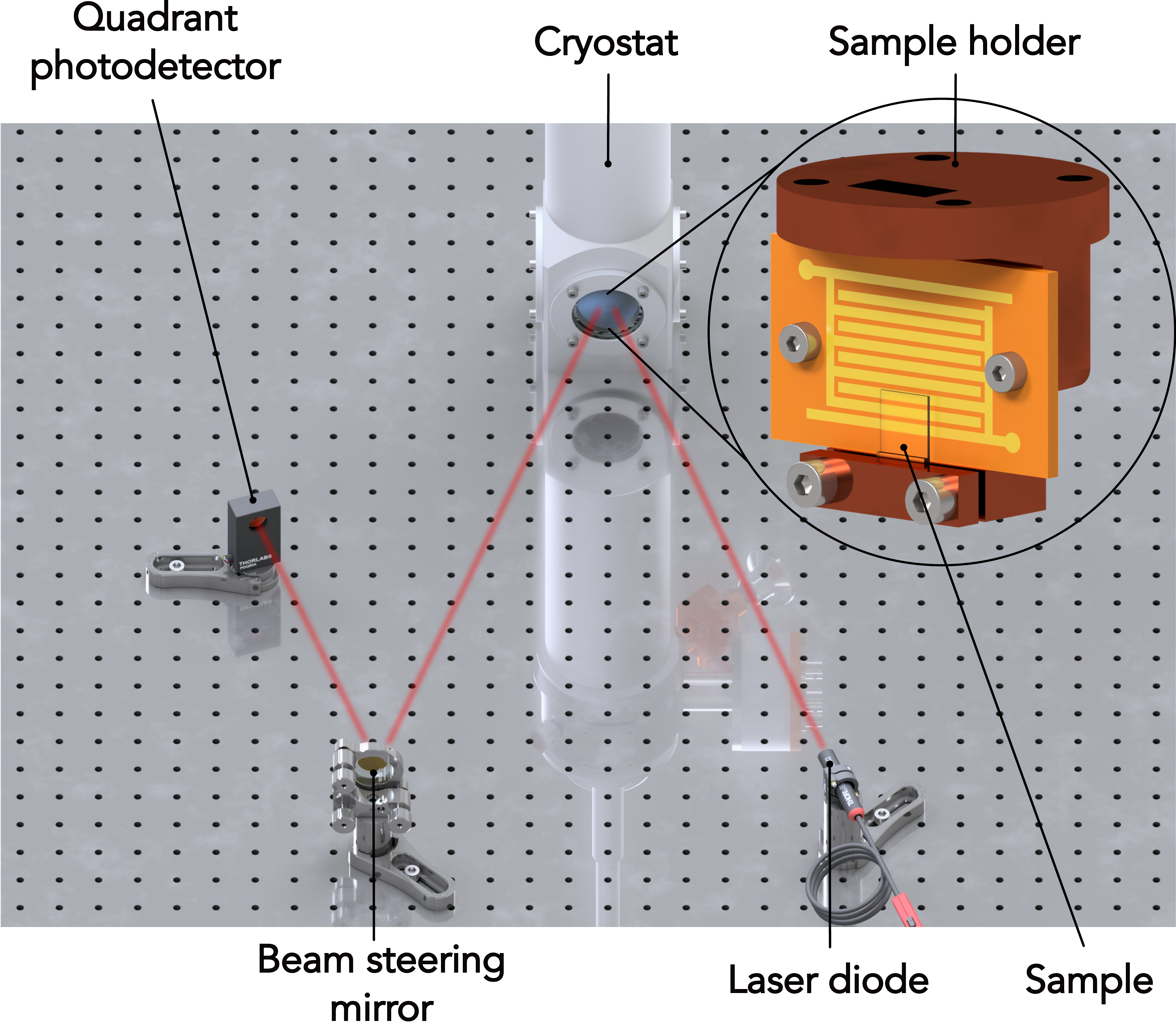}
    \caption{\label{fig:setup}Schematic of the experimental setup for mechanical loss measurements. The laser beam is reflected by the surface of the sample and the oscillation is detected by a quadrant photodetector.}
\end{figure}

Figure~\ref{fig:setup} depicts the schematic of our setup, constructed on an optical table to mitigate the impact of seismic disturbances.
The sample is mounted in a Lake Shore SuperTran ST-100 continuous flow cryostat, with pressures ranging from ${p_{300\mathrm{K}} = \SI{2e-6}{\milli\bar}}$ to ${p_{15\mathrm{K}} = \SI{3e-7}{\milli\bar}}$, achieved using a rotary vane pump and a turbomolecular pump.
Using liquid helium, the lowest achievable temperature is \SI{15}{\K}, measured at the end of the sample holder next to the sample.

The sample's excitation is controlled by a piezoelectric actuator clamped between the sample and the clamp, and driven by a frequency generator.
Although the use of a comb structure with a highly amplified voltage, as shown in Fig.~\ref{fig:setup}, was not used for the Eu:YSO samples, it is still included because it went into the COMSOL Multiphysics\textsuperscript{\textregistered} simulations discussed below.~\cite{Comsol}

The sample's oscillation is detected using the optical lever readout technique:
A laser beam from a \SI{635}{\nano\meter}/\SI{1}{\milli\watt} laser diode is reflected from the surface of the sample back into the center of a quadrant photodetector.
The quadrant photodetector containing four separate areas allows the detection of any sample motion in both horizontal and vertical directions.
The reflected beam provides information about the frequency of the vibration and its amplitude corresponding to its strength.

\section{Results}
\begin{figure}[tb]
    \centering
    \includegraphics[width=\linewidth]{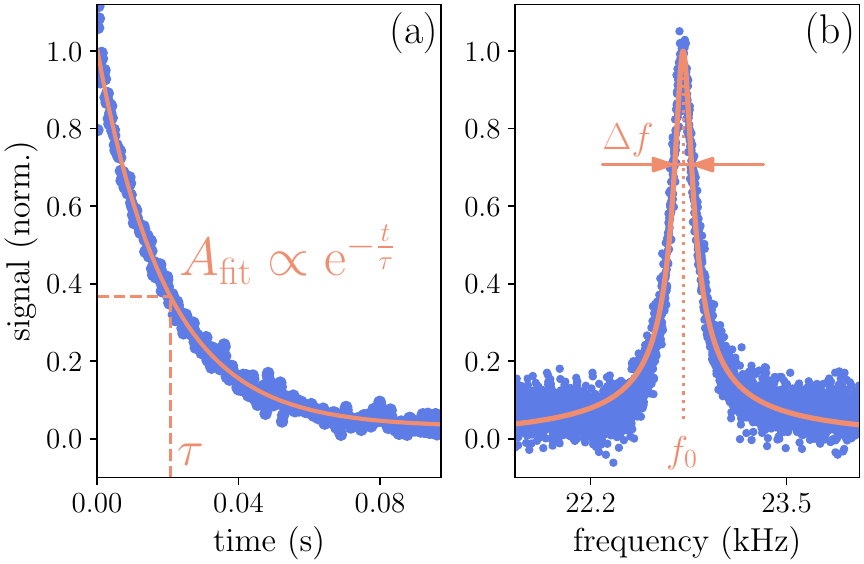}
    \caption{\label{fig:ringdown_lorentzian}Determination of the mechanical quality factor.
    a) The time constant $\tau$ can be obtained by fitting the measured ringdown (blue dots) with an exponential decaying function as $A_{\mathrm{fit}} \propto \mathrm{e}^{-t/\tau}$.
    b) In some cases where the $Q$ factor is too low to obtain a suitable ringdown, the $Q$ factor can be calculated using the following relation $Q=f_0/\Delta f$.}
\end{figure}

The key mechanism to determine the loss angle of a given material is the mechanical ringdown.
Whenever an arbitrary sample is  excited at one of its mechanical resonance frequencies, the decaying oscillation signal after switching of the excitation can be described as that of a damped harmonic oscillator.~\cite{Aspelmeyer2014}
The corresponding oscillation amplitude is given by
\begin{equation}
    A(t) = A_0 \, \mathrm{e}^{-t/\tau} \cos{(\omega t+\phi_0)} \, ,
\end{equation}
where $A_0$ represents the initial amplitude and $\phi_0$ is an arbitrary phase offset. 
The characteristic time constant, $\tau$, denotes the time for the amplitude to decay to a factor of $1/\mathrm{e}$ of its initial value as shown in Fig.~\ref{fig:ringdown_lorentzian}(a). 
For the assumption of weak damping, the quality factor $Q$ can be directly calculated with the resonant frequency $f_0$ and the time constant $\tau$~\cite{Nowick1972}
\begin{equation}
    Q = \pi f_0 \tau \, .
\end{equation}
Furthermore, in cases of high damping resulting in a low quality factor, the quality factor can be computed using the relation between the resonant frequency $f_0$ and the resonance width $\Delta f$ as shown in Fig.~\ref{fig:ringdown_lorentzian}(b)
\begin{equation}\label{eq:deltaF}
    Q = \frac{f_0}{\Delta f} \, .
\end{equation}
Finally, the mechanical loss factor is expressed as the inverse of the mechanical quality factor~\cite{Rowan1997}
\begin{equation}
    \phi (f_0) = \frac{1}{Q} \, .
\end{equation}
It is important to note that the loss factor generally is frequency-dependent.

\begin{figure}[tb]
    \centering
    \includegraphics[width=\linewidth]{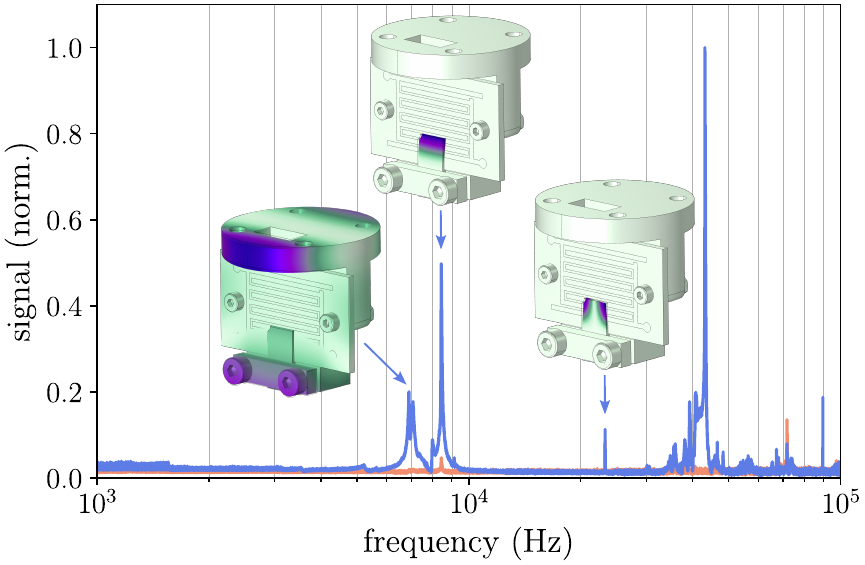}
    \caption{\label{fig:frequency_spectrum}Frequency spectrum of sample A ranging from \SI{1}{\kilo\Hz} to \SI{100}{\kilo\Hz} at $T=\SI{80}{\K}$. 
    The blue line represents the signal of the longitudinal axis, while the orange line represents the signal of the lateral axis. 
    COMSOL simulations are carried out to see which peak corresponds to an oscillation of the sample.}
\end{figure}

In Fig.~\ref{fig:frequency_spectrum}, an exemplary frequency spectrum of sample A is given at a temperature of \SI{80}{\K}.
Here, the signal obtained from the quadrant photodetector is split into two components, corresponding to oscillations from either flexural or torsional modes.
To avoid mistakenly evaluating the sample holder's mechanical resonance, COMSOL simulations are carried out to identify the sample's resonances among the measured peaks.
Unfortunately, due to the thickness of oscillators B and C, it was only possible to excite the first flexural mode in order to obtain a suitable signal-to-noise ratio (SNR) for the readout process. 
In addition, a reduction in temperature leads to a reduction in the excitation amplitude of the piezoelectric actuator additionally challenging the sample excitation.

The Young's modulus is a temperature-dependent material property and determines the resonance frequency. Thus, going to cryogenic temperatures requires tracking the resonance frequency for the loss measurements. 
Specifically, the temperature dependence of the Young's modulus $E$ can be estimated as~\cite{Gysin2004}
\begin{equation}\label{eq:young}
    E(T) = E_0 - B T \mathrm{e}^{-T_0/T} \, ,
\end{equation}
where $E_0$ denotes the Young's modulus at \SI{0}{\K}, $B$ is a temperature-independent constant related to its bulk modulus and $T_0$ is related to the Debye temperature.
Since the frequency is connected to the Young's modulus via $f\propto\sqrt{E}$~\cite{Reid2006,Rast2000}, the experimental data of the resonance frequency can be fitted with the temperature dependence of Eq.~(\ref{eq:young}) to calculate the temperature-dependent Young's modulus, starting from its room temperature value of \SI{135}{\giga\Pa}.~\cite{Motte2019}

\begin{figure}[tb]
    \centering
    \includegraphics[width=\linewidth]{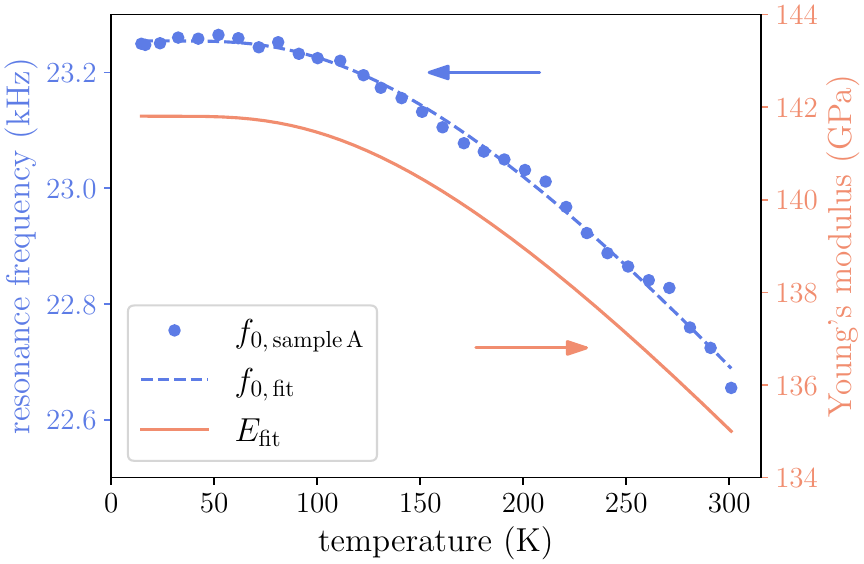}
    \caption{\label{fig:freq_shift}Shift of the resonance frequency for the torsional mode of sample A (blue dots). The corresponding shift of the Young's modulus (orange line) can be obtained from the frequency fit (dashed blue line).}
\end{figure}

In Fig.~\ref{fig:freq_shift}, we present the frequency shift of the torsional mode of sample A as well as the shift of the Young's modulus.
Here, the data of sample A is used as with this sample we were able to achieve the lowest temperature of \SI{15}{\K}.
The relative frequency shifts of the other samples/modes are at the same percentage level as the data shown in Fig.~\ref{fig:freq_shift} until the SNR gets too high.
From the fitted Young's modulus, we can derive the zero-Kelvin value of the Young's modulus, $E_0 = \SI{141.8}{\giga\pascal}$, since the resonance frequency remains constant as the temperature approaches \SI{0}{\kelvin}.

Using these mechanical eigenmodes, we perform mechanical ringdown measurements by exciting the sample and measuring the decaying signal of the mechanical oscillation as soon as the excitation is switched off.
As already mentioned, due to the thickness of samples B and C and the temperature-dependent excitation amplitude of the piezoelectric actuator, it was not possible to measure the loss down to \SI{15}{\K}.
For sample B, the lowest temperature is \SI{80}{\K}, whereas for sample C, it is \SI{180}{\K}.

\begin{figure}[tb]
    \centering
    \includegraphics[width=\linewidth]{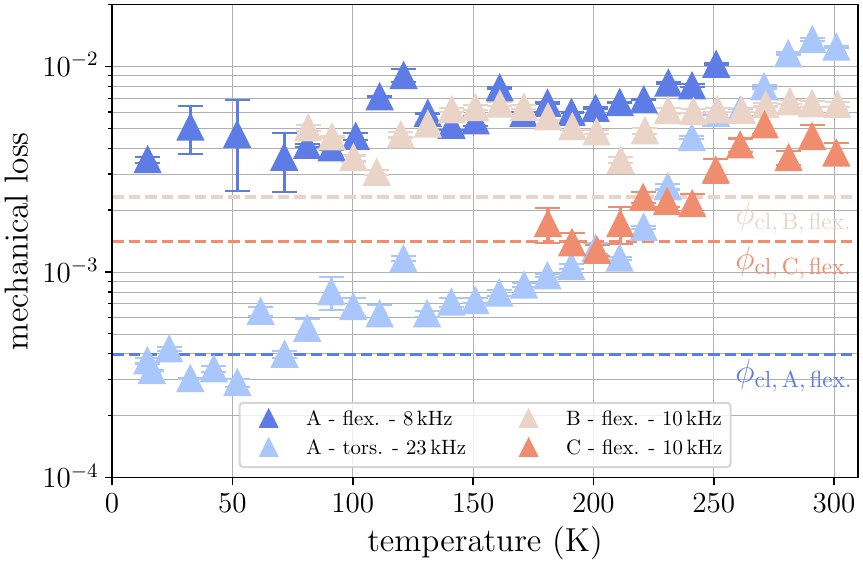}
    \caption{\label{fig:losses_results}Temperature-dependent mechanical losses for sample A, B, and C. The dashed lines correspond to the clamping losses of the flexural mode and show therefore the limitation of the loss angle of sample B and C. The lowest measured loss is $\phi=\num{2.72e-4}$ at $T=\SI{52}{\K}$.}
\end{figure}

Figure~\ref{fig:losses_results} shows the temperature-dependent mechanical losses for all three samples.
First, thermoelastic damping (TED) is evaluated as a possible loss mechanism.
During each oscillation period of a vibration mode, there are parts of the samples that are compressed and parts that are stretched.~\cite{Lifshitz1999}
This results in a temperature gradient between these zones and lead to an irreversible heat flux where energy is dissipated.~\cite{Fejer2004}
Since the literature does not yet provide temperature-dependent material properties for Eu:YSO, TED is calculated for room temperature to determine whether the loss angle is limited by TED.
Generally, the quality factor is defined as 
\begin{equation}
    Q = 2\pi\frac{E}{\Delta E} \, ,
\end{equation}
where $E$ is the total energy and $\Delta E$ the energy that is dissipated within an oscillation period.~\cite{Cagnoli2018}

In the case of TED, the dissipated energy from the heat flux can be derived from the heat equation, where the energy loss per cycle is caused by an entropy rise. In general, TED of cantilever structures can also be calculated analytically, however, this is contingent upon the assumption that the length is much greater than the width of the beam.~\cite{Lifshitz1999} A numerical expression that can be used for finite element simulations is given by Heinert \textit{et al.}~\cite{Heinert2010}
\begin{equation}
    \Delta E = \pi \int_V \alpha_{ij} \hat{\sigma}_{ij} \Im{(\hat{T})} \, \mathrm{d}V \, ,
\end{equation}
where $\alpha_{ij}$ is the tensor of thermal expansion and $\hat{\sigma}_{ij}$ is the stress tensor for a fully anisotropic treatment.
The imaginary part of the complex temperature field $\Im{(\hat{T})}$ is taken from the solution from the heat equation.
We compute TED numerically via COMSOL with the material constants given in Tab.~\ref{tab:material_parameters}.

For sample A, the TED for the flexural mode is ${\phi_{\mathrm{TED,A,1}}=\num{2.21e-5}}$ and for the (predominantly) torsional mode ${\phi_{\mathrm{TED,A,2}}=\num{6.83e-6}}$.
The TED of the flexural mode of sample B and C is ${\phi_{\mathrm{TED,B,1}}=\num{1.02e-6}}$ and ${\phi_{\mathrm{TED,C,1}}=\num{1.53e-6}}$, respectively.
Since all these values are much smaller then all the measured losses, we can conclude that TED is not a limiting factor of our measurements.
TED typically decreases towards lower temperatures, and the calculated TED values at room temperature are not even reached for the lowest measured loss.
Consequently, to understand the difference in mechanical loss behavior between the different samples, a more detailed examination of the oscillators' geometry and the associated clamping is required.

The mechanical quality factor of the clamping part can be analytically expressed as~\cite{Hao2003}
\begin{equation}\label{eq:qcl}
    Q_{\mathrm{cl}} = \left( \frac{0.24(1-\sigma)}{(1+\sigma)\psi} \right) \frac{1}{(\beta_n \chi_n)^2} \left( \frac{L}{t} \right)^3 \, ,
\end{equation}
where $\sigma$ is the Poisson's ratio, and $\beta_n$ and $\chi_n$ are the mode constants and mode shape factors, respectively, which are related to the $n$th mode of a clamped-free beam resonator.
Both constants are derived from the beam theory of flexural vibrations.
The parameter $\psi$ is defined as
\begin{equation}
    \psi = \int_0^{\infty} \frac{\sqrt{\zeta^2-\mu^2}}{\left(2\zeta^2-\mu^2\right)^2-4\zeta^2\sqrt{\zeta^2-\mu^2}\sqrt{\zeta^2-1}} \, \mathrm{d}\zeta \, ,
\end{equation}
with $\zeta = \xi c_{\mathrm{L}}/\omega$ and $\mu=c_{\mathrm{L}}/c_{\mathrm{T}}$ representing the ratio of the propagation velocities of the longitudinal and transverse waves.
The parameter $\xi$ results from the Fourier transformation of the elastic wave equation.~\cite{Hao2003}

The prefactor of Eq.~(\ref{eq:qcl}) with the constants for the first flexural mode as $\beta_1=\num{0.597}$ and $\chi_1=\num{-0.734}$ indicates that the quality factor of the clamping in this case is $Q_{\mathrm{cl}}\approx 2 (L/t)^3$, depending solely on the length-to-thickness ratio of the oscillator.
This simplified formula only holds for the first fundamental mode, as the mode constants $\beta_n$ and $\chi_n$ are larger for higher-order modes~\cite{Hao2003}, resulting in lower support loss.
Therefore, a beam resonator with a low length-to-thickness ratio will be limited more severely by clamping losses than a resonator with a high ratio.
Generally, clamping losses are temperature-independent.~\cite{Hosaka1995, Jimbo1968}

The clamping losses of the utilized samples are calculated for the first flexural mode and highlighted as a dashed line in Fig.~\ref{fig:losses_results} in order to check if the mechanical loss of the oscillator is being limited by the clamping effect.~\cite{Czaplewski2005}
Note that Eq.~(\ref{eq:qcl}) is only valid for flexural vibrations.
For sample A, there is no clamp related limitation of the losses—neither for the flexural mode nor the torsional mode.
Due to the different geometry which are limited in turn by the thickness of the available crystals, sample B and C, however, are limited by clamping losses. 
This indicates the minimum measured losses of sample A are closer to the intrinsic material losses relevant for the noise estimation compared to samples B and C. 

Moreover, the torsional mode of sample A exhibits much lower losses at cryogenic temperatures compared to the flexural mode.
A minimum loss angle of ${\phi=\num{2.72e-4}}$ for the torsional mode was measured at $T=\SI{52}{\K}$, corresponding to a quality factor of $Q = 3676$.
The different loss behavior of the torsional mode compared to the flexural mode can be attributed to the anisotropic crystal structure.
Since sample C is oriented $\SI{90}{\degree}$ relative to sample A, they share the same neutral axis when sample A oscillates in the torsional mode.
This could lead to a different mechanical behavior, as observed in other crystalline materials.~\cite{Nawrodt2008a}
Such a difference is consistent with the loss behavior of sample C, as the trend to lower temperatures is similar to that of the torsional mode.
Furthermore, we stress that this loss value is a worst-case scenario for the noise performance and thus the fractional frequency stability.
Typically, for samples with lower surface-to-volume ratio lower mechanical losses can be expected.~\cite{Gretarsson1999, Ageev2004, Penn2006, Nawrodt2013} 

\begin{table}[tb]
    \caption{\label{tab:material_parameters}Material parameters used for TED simulation.}
    \begin{tabular}{cc}
        \hline \hline
        Parameter & Value \cite{Motte2019,Sun2008,Marion1990} \\
        \hline
        Young's modulus $E$ & $\SI{135}{\giga\Pa}$ \\
        Poisson's ratio $\sigma$ & $\num{0.31}$ \\
        Coefficient of thermal expansion $\alpha$ & $\SI{7.4e-6}{\per\K}$ \\
        Thermal conductivity $\kappa$ & $\SI{4.49}{\watt\per\meter\per\K}$ \\
        Specific heat capacity $c_p$ & $\SI{646}{\J\per\kg\per\K}$ \\
        Mass density $\rho$ & $\SI{4440}{\kg\per\meter\cubed}$ \\
        \hline \hline
    \end{tabular}
\end{table}

Using the analytical expression from Hartman \textit{et al.}~\cite{Hartman2024} to calculate the power spectral density we are able to provide an upper limit for fractional frequency stability $\delta\nu/\nu_0$ based on our measured mechanical loss data.
The power spectral density for a cubic crystal of side $l$ is given as
\begin{equation}\label{eq:thermalnoise}
S_{\delta \nu/\nu_0} (f) = \frac{2k_{\mathrm{B}}T}{\pi f} \frac{E_0 \phi}{(1-2\sigma) l^3} \sum_i k_i^2, 
\end{equation}
where $k_{\mathrm{B}}$ is the Boltzmann's constant, $k_i$ represents the stress/frequency-strain modulus measured along the $i$th dielectric axis scaled to fractional frequency for the substitution site under investigation.~\cite{Galland2020, Hartman2024}
The other parameters are given by Mirzai \textit{et al.}~\cite{Mirzai2021}.

The relevant loss values used for the noise calculation are those of the torsional mode of sample A, as these are closer to the intrinsic loss value of the material.
Although SHB laser stabilization typically occurs at \SI{4}{\K} or even lower due to the lifetime and other spectroscopic properties~\cite{Cook2015, Lin2024}, it is reasonable to assume that the lowest measured loss angle measured down to \SI{15}{\K} remains valid at these temperatures at least, or gives an absolute upper bound.
We can thus estimate an Allan deviation of $\sigma_{\delta \nu/\nu_0} = \num{9.2e-18}$ in a \SI{1}{\centi\meter\cubed} cubic crystal for site 1, which is the substitution site that is metrologically relevant due to its stronger absorption.~\cite{Yano1991,Yano1992}
This limit can be further lowered to $\sigma_{\delta \nu/\nu_0} = \num{2.5e-18}$ if the SHB experiments are operated at \SI{300}{\milli\kelvin}, which also efficiently suppresses the temperature fluctuation induced frequency noise.~\cite{Lin2024}
This result for the relative frequency stability surpasses the performance of the world's most frequency-stable cavity-referenced lasers.~\cite{Matei2017}

\section{Conclusion}
In conclusion, we present mechanical loss measurements from room temperature down to \SI{15}{\K} for Eu$^{3+}$:Y$_2$SiO$_5$ (Eu:YSO)—a promising candidate for the next generation of ultra-stable lasers. 
Our results show a maximum quality factor of $Q=3676$, corresponding to a loss angle of $\phi=\num{2.72e-4}$ at \SI{52}{\K}, which is much higher than the value of $Q=\num{2500}$ measured by Ohta \textit{et al.}~\cite{Ohta2021} at \SI{4}{\K}.
Consequently, the avenue of frequency stabilization through the SHB technique emerges as a highly promising path for the next generation of ultra-stable lasers.

This project (20FUN08 NEXTLASERS) has received funding from the EMPIR programme co-financed by the Participating States and from the European Union’s Horizon 2020 research and innovation programme, and by the Deutsche Forschungsgemeinschaft (DFG, German Research Foundation) under Germany’s Excellence Strategy–EXC-2123 QuantumFrontiers–390837967.

\section*{Author declarations}
\subsection*{Conflict of Interest}
The authors have no conflicts to disclose.

\subsection*{Author Contributions}
\textbf{Nico Wagner:} Conceptualization (lead); Formal analysis (lead); Investigation (lead); Visualization (lead); Writing – original draft (lead); Writing – review \& editing (lead). \textbf{Johannes Dickmann:} Conceptualization (supporting); Formal analysis (supporting); Investigation (supporting); Visualization (supporting); Writing – review \& editing (supporting). \textbf{Bess Fang:} Conceptualization (equal); Formal analysis (equal); Investigation (equal); Writing – original draft (equal); Writing – review \& editing (equal). \textbf{Michael T. Hartman:} Conceptualization (equal); Formal analysis (equal); Investigation (equal); Writing – original draft (equal); Writing – review \& editing (equal). \textbf{Stefanie Kroker:} Conceptualization (lead); Formal analysis (lead); Investigation (lead); Funding acquisition (lead); Methodology (lead); Resources (lead); Supervision (lead); Writing – review \& editing (lead).

\section*{Data availability}
The data that support the findings of this study are available from the corresponding author upon reasonable request.

\bibliography{bibliography}

\providecommand{\noopsort}[1]{}\providecommand{\singleletter}[1]{#1}%
\begin{thebibliography}{52}%
\makeatletter
\providecommand \@ifxundefined [1]{%
 \@ifx{#1\undefined}
}%
\providecommand \@ifnum [1]{%
 \ifnum #1\expandafter \@firstoftwo
 \else \expandafter \@secondoftwo
 \fi
}%
\providecommand \@ifx [1]{%
 \ifx #1\expandafter \@firstoftwo
 \else \expandafter \@secondoftwo
 \fi
}%
\providecommand \natexlab [1]{#1}%
\providecommand \enquote  [1]{``#1''}%
\providecommand \bibnamefont  [1]{#1}%
\providecommand \bibfnamefont [1]{#1}%
\providecommand \citenamefont [1]{#1}%
\providecommand \href@noop [0]{\@secondoftwo}%
\providecommand \href [0]{\begingroup \@sanitize@url \@href}%
\providecommand \@href[1]{\@@startlink{#1}\@@href}%
\providecommand \@@href[1]{\endgroup#1\@@endlink}%
\providecommand \@sanitize@url [0]{\catcode `\\12\catcode `\$12\catcode
  `\&12\catcode `\#12\catcode `\^12\catcode `\_12\catcode `\%12\relax}%
\providecommand \@@startlink[1]{}%
\providecommand \@@endlink[0]{}%
\providecommand \url  [0]{\begingroup\@sanitize@url \@url }%
\providecommand \@url [1]{\endgroup\@href {#1}{\urlprefix }}%
\providecommand \urlprefix  [0]{URL }%
\providecommand \Eprint [0]{\href }%
\providecommand \doibase [0]{http://dx.doi.org/}%
\providecommand \selectlanguage [0]{\@gobble}%
\providecommand \bibinfo  [0]{\@secondoftwo}%
\providecommand \bibfield  [0]{\@secondoftwo}%
\providecommand \translation [1]{[#1]}%
\providecommand \BibitemOpen [0]{}%
\providecommand \bibitemStop [0]{}%
\providecommand \bibitemNoStop [0]{.\EOS\space}%
\providecommand \EOS [0]{\spacefactor3000\relax}%
\providecommand \BibitemShut  [1]{\csname bibitem#1\endcsname}%
\let\auto@bib@innerbib\@empty
\bibitem [{\citenamefont {Schioppo}\ \emph {et~al.}(2017)\citenamefont
  {Schioppo}, \citenamefont {Brown}, \citenamefont {McGrew}, \citenamefont
  {Hinkley}, \citenamefont {Fasano}, \citenamefont {Beloy}, \citenamefont
  {Yoon}, \citenamefont {Milani}, \citenamefont {Nicolodi}, \citenamefont
  {Sherman}, \citenamefont {Phillips}, \citenamefont {Oates},\ and\
  \citenamefont {Ludlow}}]{schioppo2017ultrastable}%
  \BibitemOpen
  \bibfield  {author} {\bibinfo {author} {\bibfnamefont {M.}~\bibnamefont
  {Schioppo}}, \bibinfo {author} {\bibfnamefont {R.~C.}\ \bibnamefont {Brown}},
  \bibinfo {author} {\bibfnamefont {W.~F.}\ \bibnamefont {McGrew}}, \bibinfo
  {author} {\bibfnamefont {N.}~\bibnamefont {Hinkley}}, \bibinfo {author}
  {\bibfnamefont {R.~J.}\ \bibnamefont {Fasano}}, \bibinfo {author}
  {\bibfnamefont {K.}~\bibnamefont {Beloy}}, \bibinfo {author} {\bibfnamefont
  {T.~H.}\ \bibnamefont {Yoon}}, \bibinfo {author} {\bibfnamefont
  {G.}~\bibnamefont {Milani}}, \bibinfo {author} {\bibfnamefont
  {D.}~\bibnamefont {Nicolodi}}, \bibinfo {author} {\bibfnamefont {J.~A.}\
  \bibnamefont {Sherman}}, \bibinfo {author} {\bibfnamefont {N.~B.}\
  \bibnamefont {Phillips}}, \bibinfo {author} {\bibfnamefont {C.~W.}\
  \bibnamefont {Oates}}, \ and\ \bibinfo {author} {\bibfnamefont {A.~D.}\
  \bibnamefont {Ludlow}},\ }\bibfield  {title} {\enquote {\bibinfo {title}
  {Ultrastable optical clock with two cold-atom ensembles},}\ }\href@noop {}
  {\bibfield  {journal} {\bibinfo  {journal} {Nat. Photon.}\ }\textbf {\bibinfo
  {volume} {11}},\ \bibinfo {pages} {48--52} (\bibinfo {year}
  {2017})}\BibitemShut {NoStop}%
\bibitem [{\citenamefont {Tyumenev}\ \emph {et~al.}(2016)\citenamefont
  {Tyumenev}, \citenamefont {Favier}, \citenamefont {Bilicki}, \citenamefont
  {Bookjans}, \citenamefont {Targat}, \citenamefont {Lodewyck}, \citenamefont
  {Nicolodi}, \citenamefont {Coq}, \citenamefont {Abgrall}, \citenamefont
  {Guéna}, \citenamefont {Sarlo},\ and\ \citenamefont
  {Bize}}]{tyumenev2016comparing}%
  \BibitemOpen
  \bibfield  {author} {\bibinfo {author} {\bibfnamefont {R.}~\bibnamefont
  {Tyumenev}}, \bibinfo {author} {\bibfnamefont {M.}~\bibnamefont {Favier}},
  \bibinfo {author} {\bibfnamefont {S.}~\bibnamefont {Bilicki}}, \bibinfo
  {author} {\bibfnamefont {E.}~\bibnamefont {Bookjans}}, \bibinfo {author}
  {\bibfnamefont {R.~L.}\ \bibnamefont {Targat}}, \bibinfo {author}
  {\bibfnamefont {J.}~\bibnamefont {Lodewyck}}, \bibinfo {author}
  {\bibfnamefont {D.}~\bibnamefont {Nicolodi}}, \bibinfo {author}
  {\bibfnamefont {Y.~L.}\ \bibnamefont {Coq}}, \bibinfo {author} {\bibfnamefont
  {M.}~\bibnamefont {Abgrall}}, \bibinfo {author} {\bibfnamefont
  {J.}~\bibnamefont {Guéna}}, \bibinfo {author} {\bibfnamefont {L.~D.}\
  \bibnamefont {Sarlo}}, \ and\ \bibinfo {author} {\bibfnamefont
  {S.}~\bibnamefont {Bize}},\ }\bibfield  {title} {\enquote {\bibinfo {title}
  {Comparing a mercury optical lattice clock with microwave and optical
  frequency standards},}\ }\href@noop {} {\bibfield  {journal} {\bibinfo
  {journal} {New J. Phys.}\ }\textbf {\bibinfo {volume} {18}},\ \bibinfo
  {pages} {113002} (\bibinfo {year} {2016})}\BibitemShut {NoStop}%
\bibitem [{\citenamefont {Lisdat}\ \emph {et~al.}(2016)\citenamefont {Lisdat},
  \citenamefont {Grosche}, \citenamefont {Quintin} \emph
  {et~al.}}]{lisdat2016clock}%
  \BibitemOpen
  \bibfield  {author} {\bibinfo {author} {\bibfnamefont {C.}~\bibnamefont
  {Lisdat}}, \bibinfo {author} {\bibfnamefont {G.}~\bibnamefont {Grosche}},
  \bibinfo {author} {\bibfnamefont {N.}~\bibnamefont {Quintin}},  \emph
  {et~al.},\ }\bibfield  {title} {\enquote {\bibinfo {title} {A clock network
  for geodesy and fundamental science},}\ }\href@noop {} {\bibfield  {journal}
  {\bibinfo  {journal} {Nat. Commun.}\ }\textbf {\bibinfo {volume} {7}},\
  \bibinfo {pages} {12443} (\bibinfo {year} {2016})}\BibitemShut {NoStop}%
\bibitem [{\citenamefont {Ushijima}\ \emph {et~al.}(2015)\citenamefont
  {Ushijima}, \citenamefont {Takamoto}, \citenamefont {Das}, \citenamefont
  {Ohkubo},\ and\ \citenamefont {Katori}}]{ushijima2015cryogenic}%
  \BibitemOpen
  \bibfield  {author} {\bibinfo {author} {\bibfnamefont {I.}~\bibnamefont
  {Ushijima}}, \bibinfo {author} {\bibfnamefont {M.}~\bibnamefont {Takamoto}},
  \bibinfo {author} {\bibfnamefont {M.}~\bibnamefont {Das}}, \bibinfo {author}
  {\bibfnamefont {T.}~\bibnamefont {Ohkubo}}, \ and\ \bibinfo {author}
  {\bibfnamefont {H.}~\bibnamefont {Katori}},\ }\bibfield  {title} {\enquote
  {\bibinfo {title} {Cryogenic optical lattice clocks},}\ }\href@noop {}
  {\bibfield  {journal} {\bibinfo  {journal} {Nat. Photon.}\ }\textbf {\bibinfo
  {volume} {9}},\ \bibinfo {pages} {185--189} (\bibinfo {year}
  {2015})}\BibitemShut {NoStop}%
\bibitem [{\citenamefont {Nicholson}\ \emph {et~al.}(2015)\citenamefont
  {Nicholson} \emph {et~al.}}]{nicholson2015systematic}%
  \BibitemOpen
  \bibfield  {author} {\bibinfo {author} {\bibfnamefont {T.~L.}\ \bibnamefont
  {Nicholson}} \emph {et~al.},\ }\bibfield  {title} {\enquote {\bibinfo {title}
  {Systematic evaluation of an atomic clock at 2$\times 10^{-18}$ total
  uncertainty},}\ }\href@noop {} {\bibfield  {journal} {\bibinfo  {journal}
  {Nat. Commun.}\ }\textbf {\bibinfo {volume} {6}},\ \bibinfo {pages} {6896}
  (\bibinfo {year} {2015})}\BibitemShut {NoStop}%
\bibitem [{\citenamefont {Abbott}\ \emph {et~al.}(2016)\citenamefont {Abbott},
  \citenamefont {Abbott}, \citenamefont {Abbott} \emph
  {et~al.}}]{abbott2016observation}%
  \BibitemOpen
  \bibfield  {author} {\bibinfo {author} {\bibfnamefont {B.~P.}\ \bibnamefont
  {Abbott}}, \bibinfo {author} {\bibfnamefont {R.}~\bibnamefont {Abbott}},
  \bibinfo {author} {\bibfnamefont {T.~D.}\ \bibnamefont {Abbott}},  \emph
  {et~al.},\ }\bibfield  {title} {\enquote {\bibinfo {title} {Observation of
  gravitational waves from a binary black hole merger},}\ }\href@noop {}
  {\bibfield  {journal} {\bibinfo  {journal} {Phys. Rev. Lett.}\ }\textbf
  {\bibinfo {volume} {116}},\ \bibinfo {pages} {061102} (\bibinfo {year}
  {2016})}\BibitemShut {NoStop}%
\bibitem [{\citenamefont {Harry}\ \emph {et~al.}(2006)\citenamefont {Harry},
  \citenamefont {Armandula}, \citenamefont {Black}, \citenamefont {Crooks},
  \citenamefont {Cagnoli}, \citenamefont {Hough}, \citenamefont {Murray},
  \citenamefont {Reid}, \citenamefont {Rowan}, \citenamefont {Sneddon},
  \citenamefont {Fejer}, \citenamefont {Route},\ and\ \citenamefont
  {Penn}}]{Harry2006}%
  \BibitemOpen
  \bibfield  {author} {\bibinfo {author} {\bibfnamefont {G.~M.}\ \bibnamefont
  {Harry}}, \bibinfo {author} {\bibfnamefont {H.}~\bibnamefont {Armandula}},
  \bibinfo {author} {\bibfnamefont {E.}~\bibnamefont {Black}}, \bibinfo
  {author} {\bibfnamefont {D.~R.~M.}\ \bibnamefont {Crooks}}, \bibinfo {author}
  {\bibfnamefont {G.}~\bibnamefont {Cagnoli}}, \bibinfo {author} {\bibfnamefont
  {J.}~\bibnamefont {Hough}}, \bibinfo {author} {\bibfnamefont
  {P.}~\bibnamefont {Murray}}, \bibinfo {author} {\bibfnamefont
  {S.}~\bibnamefont {Reid}}, \bibinfo {author} {\bibfnamefont {S.}~\bibnamefont
  {Rowan}}, \bibinfo {author} {\bibfnamefont {P.}~\bibnamefont {Sneddon}},
  \bibinfo {author} {\bibfnamefont {M.~M.}\ \bibnamefont {Fejer}}, \bibinfo
  {author} {\bibfnamefont {R.}~\bibnamefont {Route}}, \ and\ \bibinfo {author}
  {\bibfnamefont {S.~D.}\ \bibnamefont {Penn}},\ }\bibfield  {title} {\enquote
  {\bibinfo {title} {Thermal noise from optical coatings in gravitational wave
  detectors},}\ }\href@noop {} {\bibfield  {journal} {\bibinfo  {journal}
  {Appl. Opt.}\ }\textbf {\bibinfo {volume} {45}},\ \bibinfo {pages} {1569}
  (\bibinfo {year} {2006})}\BibitemShut {NoStop}%
\bibitem [{\citenamefont {Hild}\ \emph {et~al.}(2011)\citenamefont {Hild},
  \citenamefont {Abernathy}, \citenamefont {Acernese} \emph
  {et~al.}}]{Hild2011}%
  \BibitemOpen
  \bibfield  {author} {\bibinfo {author} {\bibfnamefont {S.}~\bibnamefont
  {Hild}}, \bibinfo {author} {\bibfnamefont {M.}~\bibnamefont {Abernathy}},
  \bibinfo {author} {\bibfnamefont {F.}~\bibnamefont {Acernese}},  \emph
  {et~al.},\ }\bibfield  {title} {\enquote {\bibinfo {title} {Sensitivity
  studies for third-generation gravitational wave observatories},}\ }\href@noop
  {} {\bibfield  {journal} {\bibinfo  {journal} {Class. Quantum Grav.}\
  }\textbf {\bibinfo {volume} {28}},\ \bibinfo {pages} {094013} (\bibinfo
  {year} {2011})}\BibitemShut {NoStop}%
\bibitem [{\citenamefont {Adhikari}\ \emph {et~al.}(2020)\citenamefont
  {Adhikari}, \citenamefont {Arai}, \citenamefont {Brooks} \emph
  {et~al.}}]{Adhikari2020}%
  \BibitemOpen
  \bibfield  {author} {\bibinfo {author} {\bibfnamefont {R.~X.}\ \bibnamefont
  {Adhikari}}, \bibinfo {author} {\bibfnamefont {K.}~\bibnamefont {Arai}},
  \bibinfo {author} {\bibfnamefont {A.~F.}\ \bibnamefont {Brooks}},  \emph
  {et~al.},\ }\bibfield  {title} {\enquote {\bibinfo {title} {A cryogenic
  silicon interferometer for gravitational-wave detection},}\ }\href@noop {}
  {\bibfield  {journal} {\bibinfo  {journal} {Class. Quantum Grav.}\ }\textbf
  {\bibinfo {volume} {37}},\ \bibinfo {pages} {165003} (\bibinfo {year}
  {2020})}\BibitemShut {NoStop}%
\bibitem [{\citenamefont {Xie}\ \emph {et~al.}(2017)\citenamefont {Xie},
  \citenamefont {Bouchand}, \citenamefont {Nicolodi}, \citenamefont {Giunta},
  \citenamefont {Hänsel}, \citenamefont {Lezius}, \citenamefont {Joshi},
  \citenamefont {Datta}, \citenamefont {Alexandre}, \citenamefont {Lours},
  \citenamefont {Tremblin}, \citenamefont {Santarelli}, \citenamefont
  {Holzwarth},\ and\ \citenamefont {Le~Coq}}]{xie2017photonic}%
  \BibitemOpen
  \bibfield  {author} {\bibinfo {author} {\bibfnamefont {X.}~\bibnamefont
  {Xie}}, \bibinfo {author} {\bibfnamefont {R.}~\bibnamefont {Bouchand}},
  \bibinfo {author} {\bibfnamefont {D.}~\bibnamefont {Nicolodi}}, \bibinfo
  {author} {\bibfnamefont {M.}~\bibnamefont {Giunta}}, \bibinfo {author}
  {\bibfnamefont {W.}~\bibnamefont {Hänsel}}, \bibinfo {author} {\bibfnamefont
  {M.}~\bibnamefont {Lezius}}, \bibinfo {author} {\bibfnamefont
  {A.}~\bibnamefont {Joshi}}, \bibinfo {author} {\bibfnamefont
  {S.}~\bibnamefont {Datta}}, \bibinfo {author} {\bibfnamefont
  {C.}~\bibnamefont {Alexandre}}, \bibinfo {author} {\bibfnamefont
  {M.}~\bibnamefont {Lours}}, \bibinfo {author} {\bibfnamefont {P.-A.}\
  \bibnamefont {Tremblin}}, \bibinfo {author} {\bibfnamefont {G.}~\bibnamefont
  {Santarelli}}, \bibinfo {author} {\bibfnamefont {R.}~\bibnamefont
  {Holzwarth}}, \ and\ \bibinfo {author} {\bibfnamefont {Y.}~\bibnamefont
  {Le~Coq}},\ }\bibfield  {title} {\enquote {\bibinfo {title} {Photonic
  microwave signals with zeptosecond-level absolute timing noise},}\
  }\href@noop {} {\bibfield  {journal} {\bibinfo  {journal} {Nat. Photon.}\
  }\textbf {\bibinfo {volume} {11}},\ \bibinfo {pages} {44--47} (\bibinfo
  {year} {2017})}\BibitemShut {NoStop}%
\bibitem [{\citenamefont {Numata}, \citenamefont {Kemery},\ and\ \citenamefont
  {Camp}(2004)}]{Numata2004}%
  \BibitemOpen
  \bibfield  {author} {\bibinfo {author} {\bibfnamefont {K.}~\bibnamefont
  {Numata}}, \bibinfo {author} {\bibfnamefont {A.}~\bibnamefont {Kemery}}, \
  and\ \bibinfo {author} {\bibfnamefont {J.}~\bibnamefont {Camp}},\ }\bibfield
  {title} {\enquote {\bibinfo {title} {Thermal-noise limit in the frequency
  stabilization of lasers with rigid cavities},}\ }\href {\doibase
  10.1103/PhysRevLett.93.250602} {\bibfield  {journal} {\bibinfo  {journal}
  {Phys. Rev. Lett.}\ }\textbf {\bibinfo {volume} {93}},\ \bibinfo {pages}
  {250602} (\bibinfo {year} {2004})}\BibitemShut {NoStop}%
\bibitem [{\citenamefont {Häfner}\ \emph {et~al.}(2015)\citenamefont
  {Häfner}, \citenamefont {Falke}, \citenamefont {Grebing}, \citenamefont
  {Vogt}, \citenamefont {Legero}, \citenamefont {Merimaa}, \citenamefont
  {Lisdat},\ and\ \citenamefont {Sterr}}]{Haefner2015}%
  \BibitemOpen
  \bibfield  {author} {\bibinfo {author} {\bibfnamefont {S.}~\bibnamefont
  {Häfner}}, \bibinfo {author} {\bibfnamefont {S.}~\bibnamefont {Falke}},
  \bibinfo {author} {\bibfnamefont {C.}~\bibnamefont {Grebing}}, \bibinfo
  {author} {\bibfnamefont {S.}~\bibnamefont {Vogt}}, \bibinfo {author}
  {\bibfnamefont {T.}~\bibnamefont {Legero}}, \bibinfo {author} {\bibfnamefont
  {M.}~\bibnamefont {Merimaa}}, \bibinfo {author} {\bibfnamefont
  {C.}~\bibnamefont {Lisdat}}, \ and\ \bibinfo {author} {\bibfnamefont
  {U.}~\bibnamefont {Sterr}},\ }\bibfield  {title} {\enquote {\bibinfo {title}
  {$8\times10^{-17}$ fractional laser frequency instability with a long
  room-temperature cavity},}\ }\href@noop {} {\bibfield  {journal} {\bibinfo
  {journal} {Opt. Lett.}\ }\textbf {\bibinfo {volume} {40}},\ \bibinfo {pages}
  {2112} (\bibinfo {year} {2015})}\BibitemShut {NoStop}%
\bibitem [{\citenamefont {Cole}\ \emph {et~al.}(2013)\citenamefont {Cole},
  \citenamefont {Zhang}, \citenamefont {Martin}, \citenamefont {Ye},\ and\
  \citenamefont {Aspelmeyer}}]{cole2013tenfold}%
  \BibitemOpen
  \bibfield  {author} {\bibinfo {author} {\bibfnamefont {G.~D.}\ \bibnamefont
  {Cole}}, \bibinfo {author} {\bibfnamefont {W.}~\bibnamefont {Zhang}},
  \bibinfo {author} {\bibfnamefont {M.~J.}\ \bibnamefont {Martin}}, \bibinfo
  {author} {\bibfnamefont {J.}~\bibnamefont {Ye}}, \ and\ \bibinfo {author}
  {\bibfnamefont {M.}~\bibnamefont {Aspelmeyer}},\ }\bibfield  {title}
  {\enquote {\bibinfo {title} {Tenfold reduction of {B}rownian noise in
  high-reflectivity optical coatings},}\ }\href@noop {} {\bibfield  {journal}
  {\bibinfo  {journal} {Nat. Photon.}\ }\textbf {\bibinfo {volume} {7}},\
  \bibinfo {pages} {644--650} (\bibinfo {year} {2013})}\BibitemShut {NoStop}%
\bibitem [{\citenamefont {Kedar}\ \emph {et~al.}(2023)\citenamefont {Kedar},
  \citenamefont {Yu}, \citenamefont {Oelker}, \citenamefont {Staron},
  \citenamefont {Milner}, \citenamefont {Robinson}, \citenamefont {Legero},
  \citenamefont {Riehle}, \citenamefont {Sterr},\ and\ \citenamefont
  {Ye}}]{Kedar2023}%
  \BibitemOpen
  \bibfield  {author} {\bibinfo {author} {\bibfnamefont {D.}~\bibnamefont
  {Kedar}}, \bibinfo {author} {\bibfnamefont {J.}~\bibnamefont {Yu}}, \bibinfo
  {author} {\bibfnamefont {E.}~\bibnamefont {Oelker}}, \bibinfo {author}
  {\bibfnamefont {A.}~\bibnamefont {Staron}}, \bibinfo {author} {\bibfnamefont
  {W.~R.}\ \bibnamefont {Milner}}, \bibinfo {author} {\bibfnamefont {J.~M.}\
  \bibnamefont {Robinson}}, \bibinfo {author} {\bibfnamefont {T.}~\bibnamefont
  {Legero}}, \bibinfo {author} {\bibfnamefont {F.}~\bibnamefont {Riehle}},
  \bibinfo {author} {\bibfnamefont {U.}~\bibnamefont {Sterr}}, \ and\ \bibinfo
  {author} {\bibfnamefont {J.}~\bibnamefont {Ye}},\ }\bibfield  {title}
  {\enquote {\bibinfo {title} {Frequency stability of cryogenic silicon
  cavities with semiconductor crystalline coatings},}\ }\href@noop {}
  {\bibfield  {journal} {\bibinfo  {journal} {Optica}\ }\textbf {\bibinfo
  {volume} {10}},\ \bibinfo {pages} {464} (\bibinfo {year} {2023})}\BibitemShut
  {NoStop}%
\bibitem [{\citenamefont {Dickmann}\ \emph {et~al.}(2023)\citenamefont
  {Dickmann}, \citenamefont {Sauer}, \citenamefont {Meyer}, \citenamefont
  {Gaedtke}, \citenamefont {Siefke}, \citenamefont {Br{\"u}ckner},
  \citenamefont {Plentz},\ and\ \citenamefont
  {Kroker}}]{dickmann2023experimental}%
  \BibitemOpen
  \bibfield  {author} {\bibinfo {author} {\bibfnamefont {J.}~\bibnamefont
  {Dickmann}}, \bibinfo {author} {\bibfnamefont {S.}~\bibnamefont {Sauer}},
  \bibinfo {author} {\bibfnamefont {J.}~\bibnamefont {Meyer}}, \bibinfo
  {author} {\bibfnamefont {M.}~\bibnamefont {Gaedtke}}, \bibinfo {author}
  {\bibfnamefont {T.}~\bibnamefont {Siefke}}, \bibinfo {author} {\bibfnamefont
  {U.}~\bibnamefont {Br{\"u}ckner}}, \bibinfo {author} {\bibfnamefont
  {J.}~\bibnamefont {Plentz}}, \ and\ \bibinfo {author} {\bibfnamefont
  {S.}~\bibnamefont {Kroker}},\ }\bibfield  {title} {\enquote {\bibinfo {title}
  {Experimental realization of a 12,000-finesse laser cavity based on a
  low-noise microstructured mirror},}\ }\href@noop {} {\bibfield  {journal}
  {\bibinfo  {journal} {Commun. Phys.}\ }\textbf {\bibinfo {volume} {6}},\
  \bibinfo {pages} {16} (\bibinfo {year} {2023})}\BibitemShut {NoStop}%
\bibitem [{\citenamefont {Kroker}\ \emph {et~al.}(2017)\citenamefont {Kroker},
  \citenamefont {Dickmann}, \citenamefont {Rojas~Hurtado}, \citenamefont
  {Heinert}, \citenamefont {Nawrodt}, \citenamefont {Levin},\ and\
  \citenamefont {Vyatchanin}}]{Kroker2017}%
  \BibitemOpen
  \bibfield  {author} {\bibinfo {author} {\bibfnamefont {S.}~\bibnamefont
  {Kroker}}, \bibinfo {author} {\bibfnamefont {J.}~\bibnamefont {Dickmann}},
  \bibinfo {author} {\bibfnamefont {C.~B.}\ \bibnamefont {Rojas~Hurtado}},
  \bibinfo {author} {\bibfnamefont {D.}~\bibnamefont {Heinert}}, \bibinfo
  {author} {\bibfnamefont {R.}~\bibnamefont {Nawrodt}}, \bibinfo {author}
  {\bibfnamefont {Y.}~\bibnamefont {Levin}}, \ and\ \bibinfo {author}
  {\bibfnamefont {S.~P.}\ \bibnamefont {Vyatchanin}},\ }\bibfield  {title}
  {\enquote {\bibinfo {title} {Brownian thermal noise in functional optical
  surfaces},}\ }\href@noop {} {\bibfield  {journal} {\bibinfo  {journal} {Phys.
  Rev. D}\ }\textbf {\bibinfo {volume} {96}},\ \bibinfo {pages} {022002}
  (\bibinfo {year} {2017})}\BibitemShut {NoStop}%
\bibitem [{\citenamefont {Kessler}\ \emph {et~al.}(2012)\citenamefont
  {Kessler}, \citenamefont {Hagemann}, \citenamefont {Grebing}, \citenamefont
  {Legero}, \citenamefont {Sterr}, \citenamefont {Riehle}, \citenamefont
  {Martin}, \citenamefont {Chen},\ and\ \citenamefont {Ye}}]{kessler2012sub}%
  \BibitemOpen
  \bibfield  {author} {\bibinfo {author} {\bibfnamefont {T.}~\bibnamefont
  {Kessler}}, \bibinfo {author} {\bibfnamefont {C.}~\bibnamefont {Hagemann}},
  \bibinfo {author} {\bibfnamefont {C.}~\bibnamefont {Grebing}}, \bibinfo
  {author} {\bibfnamefont {T.}~\bibnamefont {Legero}}, \bibinfo {author}
  {\bibfnamefont {U.}~\bibnamefont {Sterr}}, \bibinfo {author} {\bibfnamefont
  {F.}~\bibnamefont {Riehle}}, \bibinfo {author} {\bibfnamefont
  {M.}~\bibnamefont {Martin}}, \bibinfo {author} {\bibfnamefont
  {L.}~\bibnamefont {Chen}}, \ and\ \bibinfo {author} {\bibfnamefont
  {J.}~\bibnamefont {Ye}},\ }\bibfield  {title} {\enquote {\bibinfo {title} {A
  sub-40-m{H}z-linewidth laser based on a silicon single-crystal optical
  cavity},}\ }\href@noop {} {\bibfield  {journal} {\bibinfo  {journal} {Nat.
  Photon.}\ }\textbf {\bibinfo {volume} {6}},\ \bibinfo {pages} {687--692}
  (\bibinfo {year} {2012})}\BibitemShut {NoStop}%
\bibitem [{\citenamefont {Thorpe}\ \emph {et~al.}(2011)\citenamefont {Thorpe},
  \citenamefont {Rippe}, \citenamefont {Fortier}, \citenamefont {Kirchner},\
  and\ \citenamefont {Rosenband}}]{thorpe2011frequency}%
  \BibitemOpen
  \bibfield  {author} {\bibinfo {author} {\bibfnamefont {M.~J.}\ \bibnamefont
  {Thorpe}}, \bibinfo {author} {\bibfnamefont {L.}~\bibnamefont {Rippe}},
  \bibinfo {author} {\bibfnamefont {T.~M.}\ \bibnamefont {Fortier}}, \bibinfo
  {author} {\bibfnamefont {M.~S.}\ \bibnamefont {Kirchner}}, \ and\ \bibinfo
  {author} {\bibfnamefont {T.}~\bibnamefont {Rosenband}},\ }\bibfield  {title}
  {\enquote {\bibinfo {title} {Frequency stabilization to $6\times 10^{-16}$
  via spectral-hole burning},}\ }\href@noop {} {\bibfield  {journal} {\bibinfo
  {journal} {Nat. Photon.}\ }\textbf {\bibinfo {volume} {5}},\ \bibinfo {pages}
  {688--693} (\bibinfo {year} {2011})}\BibitemShut {NoStop}%
\bibitem [{\citenamefont {Equall}\ \emph {et~al.}(1994)\citenamefont {Equall},
  \citenamefont {Sun}, \citenamefont {Cone},\ and\ \citenamefont
  {Macfarlane}}]{equall1994ultraslow}%
  \BibitemOpen
  \bibfield  {author} {\bibinfo {author} {\bibfnamefont {R.~W.}\ \bibnamefont
  {Equall}}, \bibinfo {author} {\bibfnamefont {Y.}~\bibnamefont {Sun}},
  \bibinfo {author} {\bibfnamefont {R.}~\bibnamefont {Cone}}, \ and\ \bibinfo
  {author} {\bibfnamefont {R.}~\bibnamefont {Macfarlane}},\ }\bibfield  {title}
  {\enquote {\bibinfo {title} {Ultraslow optical dephasing in
  {E}u$^{3+}$:{Y}$_{2}${S}i{O}$_{5}$},}\ }\href@noop {} {\bibfield  {journal}
  {\bibinfo  {journal} {Phys. Rev. Lett.}\ }\textbf {\bibinfo {volume} {72}},\
  \bibinfo {pages} {2179} (\bibinfo {year} {1994})}\BibitemShut {NoStop}%
\bibitem [{\citenamefont {Ohta}\ \emph {et~al.}(2021)\citenamefont {Ohta},
  \citenamefont {Herpin}, \citenamefont {Bastidas}, \citenamefont {Tawara},
  \citenamefont {Yamaguchi},\ and\ \citenamefont {Okamoto}}]{Ohta2021}%
  \BibitemOpen
  \bibfield  {author} {\bibinfo {author} {\bibfnamefont {R.}~\bibnamefont
  {Ohta}}, \bibinfo {author} {\bibfnamefont {L.}~\bibnamefont {Herpin}},
  \bibinfo {author} {\bibfnamefont {V.}~\bibnamefont {Bastidas}}, \bibinfo
  {author} {\bibfnamefont {T.}~\bibnamefont {Tawara}}, \bibinfo {author}
  {\bibfnamefont {H.}~\bibnamefont {Yamaguchi}}, \ and\ \bibinfo {author}
  {\bibfnamefont {H.}~\bibnamefont {Okamoto}},\ }\bibfield  {title} {\enquote
  {\bibinfo {title} {Rare-{E}arth-{M}ediated {O}ptomechanical {S}ystem in the
  {R}eversed {D}issipation {R}egime},}\ }\href@noop {} {\bibfield  {journal}
  {\bibinfo  {journal} {Phys. Rev. Lett.}\ }\textbf {\bibinfo {volume} {126}},\
  \bibinfo {pages} {047404} (\bibinfo {year} {2021})}\BibitemShut {NoStop}%
\bibitem [{\citenamefont {Motte}\ \emph {et~al.}(2019)\citenamefont {Motte},
  \citenamefont {Galland}, \citenamefont {Debray}, \citenamefont {Ferrier},
  \citenamefont {Goldner}, \citenamefont {Lu{\v{c}}i{\'{c}}}, \citenamefont
  {Zhang}, \citenamefont {Fang}, \citenamefont {Coq},\ and\ \citenamefont
  {Seidelin}}]{Motte2019}%
  \BibitemOpen
  \bibfield  {author} {\bibinfo {author} {\bibfnamefont {J.-F.}\ \bibnamefont
  {Motte}}, \bibinfo {author} {\bibfnamefont {N.}~\bibnamefont {Galland}},
  \bibinfo {author} {\bibfnamefont {J.}~\bibnamefont {Debray}}, \bibinfo
  {author} {\bibfnamefont {A.}~\bibnamefont {Ferrier}}, \bibinfo {author}
  {\bibfnamefont {P.}~\bibnamefont {Goldner}}, \bibinfo {author} {\bibfnamefont
  {N.}~\bibnamefont {Lu{\v{c}}i{\'{c}}}}, \bibinfo {author} {\bibfnamefont
  {S.}~\bibnamefont {Zhang}}, \bibinfo {author} {\bibfnamefont
  {B.}~\bibnamefont {Fang}}, \bibinfo {author} {\bibfnamefont {Y.~L.}\
  \bibnamefont {Coq}}, \ and\ \bibinfo {author} {\bibfnamefont
  {S.}~\bibnamefont {Seidelin}},\ }\bibfield  {title} {\enquote {\bibinfo
  {title} {Microscale {C}rystalline {R}are-{E}arth {D}oped {R}esonators for
  {S}train-{C}oupled {O}ptomechanics},}\ }\href@noop {} {\bibfield  {journal}
  {\bibinfo  {journal} {J. Mod. Phys.}\ }\textbf {\bibinfo {volume} {10}},\
  \bibinfo {pages} {1342--1352} (\bibinfo {year} {2019})}\BibitemShut {NoStop}%
\bibitem [{\citenamefont {Li}, \citenamefont {Wyon},\ and\ \citenamefont
  {Moncorge}(1992)}]{YSO_axis}%
  \BibitemOpen
  \bibfield  {author} {\bibinfo {author} {\bibfnamefont {C.}~\bibnamefont
  {Li}}, \bibinfo {author} {\bibfnamefont {C.}~\bibnamefont {Wyon}}, \ and\
  \bibinfo {author} {\bibfnamefont {R.}~\bibnamefont {Moncorge}},\ }\bibfield
  {title} {\enquote {\bibinfo {title} {Spectroscopic {P}roperties and
  {F}luorescence {D}ynamics of {E}r$^{3+}$ and {Y}b$^{3+}$ in
  {Y}$_2${S}i{O}$_5$},}\ }\href@noop {} {\bibfield  {journal} {\bibinfo
  {journal} {IEEE J. Quantum Electron.}\ }\textbf {\bibinfo {volume} {28}},\
  \bibinfo {pages} {1209--1221} (\bibinfo {year} {1992})}\BibitemShut {NoStop}%
\bibitem [{\citenamefont {Nawrodt}\ \emph {et~al.}(2013)\citenamefont {Nawrodt}
  \emph {et~al.}}]{Nawrodt2013}%
  \BibitemOpen
  \bibfield  {author} {\bibinfo {author} {\bibfnamefont {R.}~\bibnamefont
  {Nawrodt}} \emph {et~al.},\ }\bibfield  {title} {\enquote {\bibinfo {title}
  {Investigation of mechanical losses of thin silicon flexures at low
  temperatures},}\ }\href@noop {} {\bibfield  {journal} {\bibinfo  {journal}
  {Class. Quantum Grav.}\ }\textbf {\bibinfo {volume} {30}},\ \bibinfo {pages}
  {115008} (\bibinfo {year} {2013})}\BibitemShut {NoStop}%
\bibitem [{Com()}]{Comsol}%
  \BibitemOpen
  \href@noop {} {}\bibinfo {note} {COMSOL
  Multiphysics\textsuperscript{\textregistered} v.~6.0. \url{www.comsol.com}.
  COMSOL AB, Stockholm, Sweden.}\BibitemShut {Stop}%
\bibitem [{\citenamefont {Aspelmeyer}, \citenamefont {Kippenberg},\ and\
  \citenamefont {Marquardt}(2014)}]{Aspelmeyer2014}%
  \BibitemOpen
  \bibfield  {author} {\bibinfo {author} {\bibfnamefont {M.}~\bibnamefont
  {Aspelmeyer}}, \bibinfo {author} {\bibfnamefont {T.~J.}\ \bibnamefont
  {Kippenberg}}, \ and\ \bibinfo {author} {\bibfnamefont {F.}~\bibnamefont
  {Marquardt}},\ }\bibfield  {title} {\enquote {\bibinfo {title} {Cavity
  {O}ptomechanics},}\ }\href@noop {} {\bibfield  {journal} {\bibinfo  {journal}
  {Rev. Mod. Phys.}\ }\textbf {\bibinfo {volume} {86}},\ \bibinfo {pages}
  {1391--1452} (\bibinfo {year} {2014})}\BibitemShut {NoStop}%
\bibitem [{\citenamefont {Nowick}\ and\ \citenamefont
  {Berry}(1972)}]{Nowick1972}%
  \BibitemOpen
  \bibfield  {author} {\bibinfo {author} {\bibfnamefont {A.~S.}\ \bibnamefont
  {Nowick}}\ and\ \bibinfo {author} {\bibfnamefont {B.~S.}\ \bibnamefont
  {Berry}},\ }\href@noop {} {\emph {\bibinfo {title} {Anelastic relaxation in
  crystalline solids}}}\ (\bibinfo  {publisher} {Academic Press, New York},\
  \bibinfo {year} {1972})\BibitemShut {NoStop}%
\bibitem [{\citenamefont {Rowan}\ \emph {et~al.}(1997)\citenamefont {Rowan},
  \citenamefont {Hutchins}, \citenamefont {McLaren}, \citenamefont {Robertson},
  \citenamefont {Twyford},\ and\ \citenamefont {Hough}}]{Rowan1997}%
  \BibitemOpen
  \bibfield  {author} {\bibinfo {author} {\bibfnamefont {S.}~\bibnamefont
  {Rowan}}, \bibinfo {author} {\bibfnamefont {R.}~\bibnamefont {Hutchins}},
  \bibinfo {author} {\bibfnamefont {A.}~\bibnamefont {McLaren}}, \bibinfo
  {author} {\bibfnamefont {N.~A.}\ \bibnamefont {Robertson}}, \bibinfo {author}
  {\bibfnamefont {S.~M.}\ \bibnamefont {Twyford}}, \ and\ \bibinfo {author}
  {\bibfnamefont {J.}~\bibnamefont {Hough}},\ }\bibfield  {title} {\enquote
  {\bibinfo {title} {The quality factor of natural fused quartz ribbons over a
  frequency range from 6 to 160 {H}z},}\ }\href@noop {} {\bibfield  {journal}
  {\bibinfo  {journal} {Phys. Rev. A}\ }\textbf {\bibinfo {volume} {227}},\
  \bibinfo {pages} {153--158} (\bibinfo {year} {1997})}\BibitemShut {NoStop}%
\bibitem [{\citenamefont {Gysin}\ \emph {et~al.}(2004)\citenamefont {Gysin},
  \citenamefont {Rast}, \citenamefont {Ruff}, \citenamefont {Meyer},
  \citenamefont {Lee}, \citenamefont {Vettiger},\ and\ \citenamefont
  {Gerber}}]{Gysin2004}%
  \BibitemOpen
  \bibfield  {author} {\bibinfo {author} {\bibfnamefont {U.}~\bibnamefont
  {Gysin}}, \bibinfo {author} {\bibfnamefont {S.}~\bibnamefont {Rast}},
  \bibinfo {author} {\bibfnamefont {P.}~\bibnamefont {Ruff}}, \bibinfo {author}
  {\bibfnamefont {E.}~\bibnamefont {Meyer}}, \bibinfo {author} {\bibfnamefont
  {D.~W.}\ \bibnamefont {Lee}}, \bibinfo {author} {\bibfnamefont
  {P.}~\bibnamefont {Vettiger}}, \ and\ \bibinfo {author} {\bibfnamefont
  {C.}~\bibnamefont {Gerber}},\ }\bibfield  {title} {\enquote {\bibinfo {title}
  {Temperature dependence of the force sensitivity of silicon cantilevers},}\
  }\href@noop {} {\bibfield  {journal} {\bibinfo  {journal} {Phys. Rev. B}\
  }\textbf {\bibinfo {volume} {69}},\ \bibinfo {pages} {045403} (\bibinfo
  {year} {2004})}\BibitemShut {NoStop}%
\bibitem [{\citenamefont {Reid}\ \emph {et~al.}(2006)\citenamefont {Reid},
  \citenamefont {Cagnoli}, \citenamefont {Crooks}, \citenamefont {Hough},
  \citenamefont {Murray}, \citenamefont {Rowan}, \citenamefont {Fejer},
  \citenamefont {Route},\ and\ \citenamefont {Zappe}}]{Reid2006}%
  \BibitemOpen
  \bibfield  {author} {\bibinfo {author} {\bibfnamefont {S.}~\bibnamefont
  {Reid}}, \bibinfo {author} {\bibfnamefont {G.}~\bibnamefont {Cagnoli}},
  \bibinfo {author} {\bibfnamefont {D.}~\bibnamefont {Crooks}}, \bibinfo
  {author} {\bibfnamefont {J.}~\bibnamefont {Hough}}, \bibinfo {author}
  {\bibfnamefont {P.}~\bibnamefont {Murray}}, \bibinfo {author} {\bibfnamefont
  {S.}~\bibnamefont {Rowan}}, \bibinfo {author} {\bibfnamefont
  {M.}~\bibnamefont {Fejer}}, \bibinfo {author} {\bibfnamefont
  {R.}~\bibnamefont {Route}}, \ and\ \bibinfo {author} {\bibfnamefont
  {S.}~\bibnamefont {Zappe}},\ }\bibfield  {title} {\enquote {\bibinfo {title}
  {Mechanical dissipation in silicon flexures},}\ }\href@noop {} {\bibfield
  {journal} {\bibinfo  {journal} {Phys. Lett. A}\ }\textbf {\bibinfo {volume}
  {351}},\ \bibinfo {pages} {205--211} (\bibinfo {year} {2006})}\BibitemShut
  {NoStop}%
\bibitem [{\citenamefont {Rast}, \citenamefont {Wattinger},\ and\ \citenamefont
  {Meyer}(2000)}]{Rast2000}%
  \BibitemOpen
  \bibfield  {author} {\bibinfo {author} {\bibfnamefont {S.}~\bibnamefont
  {Rast}}, \bibinfo {author} {\bibfnamefont {C.}~\bibnamefont {Wattinger}}, \
  and\ \bibinfo {author} {\bibfnamefont {E.}~\bibnamefont {Meyer}},\ }\bibfield
   {title} {\enquote {\bibinfo {title} {Dynamics of damped cantilevers},}\
  }\href@noop {} {\bibfield  {journal} {\bibinfo  {journal} {Rev. Sci.
  Instrum.}\ }\textbf {\bibinfo {volume} {71}},\ \bibinfo {pages} {2772--2775}
  (\bibinfo {year} {2000})}\BibitemShut {NoStop}%
\bibitem [{\citenamefont {Lifshitz}\ and\ \citenamefont
  {Roukes}(1999)}]{Lifshitz1999}%
  \BibitemOpen
  \bibfield  {author} {\bibinfo {author} {\bibfnamefont {R.}~\bibnamefont
  {Lifshitz}}\ and\ \bibinfo {author} {\bibfnamefont {M.~L.}\ \bibnamefont
  {Roukes}},\ }\bibfield  {title} {\enquote {\bibinfo {title} {Thermoelastic
  damping in micro- and nanomechanical systems},}\ }\href@noop {} {\bibfield
  {journal} {\bibinfo  {journal} {Phys. Rev. B}\ }\textbf {\bibinfo {volume}
  {61}},\ \bibinfo {pages} {5600--5609} (\bibinfo {year} {1999})}\BibitemShut
  {NoStop}%
\bibitem [{\citenamefont {Fejer}\ \emph {et~al.}(2004)\citenamefont {Fejer},
  \citenamefont {Rowan}, \citenamefont {Cagnoli}, \citenamefont {Crooks},
  \citenamefont {Gretarsson}, \citenamefont {Harry}, \citenamefont {Hough},
  \citenamefont {Penn}, \citenamefont {Sneddon},\ and\ \citenamefont
  {Vyatchanin}}]{Fejer2004}%
  \BibitemOpen
  \bibfield  {author} {\bibinfo {author} {\bibfnamefont {M.~M.}\ \bibnamefont
  {Fejer}}, \bibinfo {author} {\bibfnamefont {S.}~\bibnamefont {Rowan}},
  \bibinfo {author} {\bibfnamefont {G.}~\bibnamefont {Cagnoli}}, \bibinfo
  {author} {\bibfnamefont {D.~R.~M.}\ \bibnamefont {Crooks}}, \bibinfo {author}
  {\bibfnamefont {A.}~\bibnamefont {Gretarsson}}, \bibinfo {author}
  {\bibfnamefont {G.~M.}\ \bibnamefont {Harry}}, \bibinfo {author}
  {\bibfnamefont {J.}~\bibnamefont {Hough}}, \bibinfo {author} {\bibfnamefont
  {S.~D.}\ \bibnamefont {Penn}}, \bibinfo {author} {\bibfnamefont {P.~H.}\
  \bibnamefont {Sneddon}}, \ and\ \bibinfo {author} {\bibfnamefont {S.~P.}\
  \bibnamefont {Vyatchanin}},\ }\bibfield  {title} {\enquote {\bibinfo {title}
  {Thermoelastic dissipation in inhomogeneous media: loss measurements and
  displacement noise in coated test masses for interferometric gravitational
  wave detectors},}\ }\href@noop {} {\bibfield  {journal} {\bibinfo  {journal}
  {Phys. Rev. D}\ }\textbf {\bibinfo {volume} {70}},\ \bibinfo {pages} {082003}
  (\bibinfo {year} {2004})}\BibitemShut {NoStop}%
\bibitem [{\citenamefont {Cagnoli}\ \emph {et~al.}(2018)\citenamefont
  {Cagnoli}, \citenamefont {Lorenzini}, \citenamefont {Cesarini}, \citenamefont
  {Piergiovanni}, \citenamefont {Granata}, \citenamefont {Heinert},
  \citenamefont {Martelli}, \citenamefont {Nawrodt}, \citenamefont {Amato},
  \citenamefont {Cassar}, \citenamefont {Dickmann}, \citenamefont {Kroker},
  \citenamefont {Lumaca}, \citenamefont {Malhaire},\ and\ \citenamefont {{Rojas
  Hurtado}}}]{Cagnoli2018}%
  \BibitemOpen
  \bibfield  {author} {\bibinfo {author} {\bibfnamefont {G.}~\bibnamefont
  {Cagnoli}}, \bibinfo {author} {\bibfnamefont {M.}~\bibnamefont {Lorenzini}},
  \bibinfo {author} {\bibfnamefont {E.}~\bibnamefont {Cesarini}}, \bibinfo
  {author} {\bibfnamefont {F.}~\bibnamefont {Piergiovanni}}, \bibinfo {author}
  {\bibfnamefont {M.}~\bibnamefont {Granata}}, \bibinfo {author} {\bibfnamefont
  {D.}~\bibnamefont {Heinert}}, \bibinfo {author} {\bibfnamefont
  {F.}~\bibnamefont {Martelli}}, \bibinfo {author} {\bibfnamefont
  {R.}~\bibnamefont {Nawrodt}}, \bibinfo {author} {\bibfnamefont
  {A.}~\bibnamefont {Amato}}, \bibinfo {author} {\bibfnamefont
  {Q.}~\bibnamefont {Cassar}}, \bibinfo {author} {\bibfnamefont
  {J.}~\bibnamefont {Dickmann}}, \bibinfo {author} {\bibfnamefont
  {S.}~\bibnamefont {Kroker}}, \bibinfo {author} {\bibfnamefont
  {D.}~\bibnamefont {Lumaca}}, \bibinfo {author} {\bibfnamefont
  {C.}~\bibnamefont {Malhaire}}, \ and\ \bibinfo {author} {\bibfnamefont
  {C.}~\bibnamefont {{Rojas Hurtado}}},\ }\bibfield  {title} {\enquote
  {\bibinfo {title} {Mode-dependent mechanical losses in disc resonators},}\
  }\href@noop {} {\bibfield  {journal} {\bibinfo  {journal} {Phys. Lett. A}\
  }\textbf {\bibinfo {volume} {382}},\ \bibinfo {pages} {2165--2173} (\bibinfo
  {year} {2018})}\BibitemShut {NoStop}%
\bibitem [{\citenamefont {Heinert}\ \emph {et~al.}(2010)\citenamefont
  {Heinert}, \citenamefont {Grib}, \citenamefont {Haughian}, \citenamefont
  {Hough}, \citenamefont {Kroker}, \citenamefont {Murray}, \citenamefont
  {Nawrodt}, \citenamefont {Rowan}, \citenamefont {Schwarz}, \citenamefont
  {Seidel},\ and\ \citenamefont {Tünnermann}}]{Heinert2010}%
  \BibitemOpen
  \bibfield  {author} {\bibinfo {author} {\bibfnamefont {D.}~\bibnamefont
  {Heinert}}, \bibinfo {author} {\bibfnamefont {A.}~\bibnamefont {Grib}},
  \bibinfo {author} {\bibfnamefont {K.}~\bibnamefont {Haughian}}, \bibinfo
  {author} {\bibfnamefont {J.}~\bibnamefont {Hough}}, \bibinfo {author}
  {\bibfnamefont {S.}~\bibnamefont {Kroker}}, \bibinfo {author} {\bibfnamefont
  {P.}~\bibnamefont {Murray}}, \bibinfo {author} {\bibfnamefont
  {R.}~\bibnamefont {Nawrodt}}, \bibinfo {author} {\bibfnamefont
  {S.}~\bibnamefont {Rowan}}, \bibinfo {author} {\bibfnamefont
  {C.}~\bibnamefont {Schwarz}}, \bibinfo {author} {\bibfnamefont
  {P.}~\bibnamefont {Seidel}}, \ and\ \bibinfo {author} {\bibfnamefont
  {A.}~\bibnamefont {Tünnermann}},\ }\bibfield  {title} {\enquote {\bibinfo
  {title} {Potential mechanical loss mechanisms in bulk materials for future
  gravitational wave detectors},}\ }\href@noop {} {\bibfield  {journal}
  {\bibinfo  {journal} {J. Phys.: Conf. Ser.}\ }\textbf {\bibinfo {volume}
  {228}},\ \bibinfo {pages} {012032} (\bibinfo {year} {2010})}\BibitemShut
  {NoStop}%
\bibitem [{\citenamefont {Hao}, \citenamefont {Erbil},\ and\ \citenamefont
  {Ayazi}(2003)}]{Hao2003}%
  \BibitemOpen
  \bibfield  {author} {\bibinfo {author} {\bibfnamefont {Z.}~\bibnamefont
  {Hao}}, \bibinfo {author} {\bibfnamefont {A.}~\bibnamefont {Erbil}}, \ and\
  \bibinfo {author} {\bibfnamefont {F.}~\bibnamefont {Ayazi}},\ }\bibfield
  {title} {\enquote {\bibinfo {title} {An analytical model for support loss in
  micromachined beam resonators with in-plane flexural vibrations},}\
  }\href@noop {} {\bibfield  {journal} {\bibinfo  {journal} {Sens. Actuator
  A-Phys.}\ }\textbf {\bibinfo {volume} {109}},\ \bibinfo {pages} {156--164}
  (\bibinfo {year} {2003})}\BibitemShut {NoStop}%
\bibitem [{\citenamefont {Hosaka}, \citenamefont {Itao},\ and\ \citenamefont
  {Kuroda}(1995)}]{Hosaka1995}%
  \BibitemOpen
  \bibfield  {author} {\bibinfo {author} {\bibfnamefont {H.}~\bibnamefont
  {Hosaka}}, \bibinfo {author} {\bibfnamefont {K.}~\bibnamefont {Itao}}, \ and\
  \bibinfo {author} {\bibfnamefont {S.}~\bibnamefont {Kuroda}},\ }\bibfield
  {title} {\enquote {\bibinfo {title} {Damping characteristics of beam-shaped
  micro-oscillators},}\ }\href@noop {} {\bibfield  {journal} {\bibinfo
  {journal} {Sens. Actuator A-Phys.}\ }\textbf {\bibinfo {volume} {49}},\
  \bibinfo {pages} {87--95} (\bibinfo {year} {1995})}\BibitemShut {NoStop}%
\bibitem [{\citenamefont {Jimbo}\ and\ \citenamefont {Itao}(1968)}]{Jimbo1968}%
  \BibitemOpen
  \bibfield  {author} {\bibinfo {author} {\bibfnamefont {Y.}~\bibnamefont
  {Jimbo}}\ and\ \bibinfo {author} {\bibfnamefont {K.}~\bibnamefont {Itao}},\
  }\bibfield  {title} {\enquote {\bibinfo {title} {Energy loss of a cantilever
  vibrator},}\ }\href@noop {} {\bibfield  {journal} {\bibinfo  {journal} {J.
  Horological Inst. Jpn.}\ }\textbf {\bibinfo {volume} {47}},\ \bibinfo {pages}
  {1--15} (\bibinfo {year} {1968})}\BibitemShut {NoStop}%
\bibitem [{\citenamefont {Czaplewski}\ \emph {et~al.}(2005)\citenamefont
  {Czaplewski}, \citenamefont {Sullivan}, \citenamefont {Friedmann},
  \citenamefont {Carr}, \citenamefont {Keeler},\ and\ \citenamefont
  {Wendt}}]{Czaplewski2005}%
  \BibitemOpen
  \bibfield  {author} {\bibinfo {author} {\bibfnamefont {D.~A.}\ \bibnamefont
  {Czaplewski}}, \bibinfo {author} {\bibfnamefont {J.~P.}\ \bibnamefont
  {Sullivan}}, \bibinfo {author} {\bibfnamefont {T.~A.}\ \bibnamefont
  {Friedmann}}, \bibinfo {author} {\bibfnamefont {D.~W.}\ \bibnamefont {Carr}},
  \bibinfo {author} {\bibfnamefont {B.~E.~N.}\ \bibnamefont {Keeler}}, \ and\
  \bibinfo {author} {\bibfnamefont {J.~R.}\ \bibnamefont {Wendt}},\ }\bibfield
  {title} {\enquote {\bibinfo {title} {Mechanical dissipation in tetrahedral
  amorphous carbon},}\ }\href@noop {} {\bibfield  {journal} {\bibinfo
  {journal} {J. Appl. Phys.}\ }\textbf {\bibinfo {volume} {97}},\ \bibinfo
  {pages} {023517} (\bibinfo {year} {2005})}\BibitemShut {NoStop}%
\bibitem [{\citenamefont {Nawrodt}\ \emph {et~al.}(2008)\citenamefont
  {Nawrodt}, \citenamefont {Zimmer}, \citenamefont {Koettig}, \citenamefont
  {Schwarz}, \citenamefont {Heinert}, \citenamefont {Hudl}, \citenamefont
  {Neubert}, \citenamefont {Thürk}, \citenamefont {Nietzsche}, \citenamefont
  {Vodel}, \citenamefont {Seidel},\ and\ \citenamefont
  {Tünnermann}}]{Nawrodt2008a}%
  \BibitemOpen
  \bibfield  {author} {\bibinfo {author} {\bibfnamefont {R.}~\bibnamefont
  {Nawrodt}}, \bibinfo {author} {\bibfnamefont {A.}~\bibnamefont {Zimmer}},
  \bibinfo {author} {\bibfnamefont {T.}~\bibnamefont {Koettig}}, \bibinfo
  {author} {\bibfnamefont {C.}~\bibnamefont {Schwarz}}, \bibinfo {author}
  {\bibfnamefont {D.}~\bibnamefont {Heinert}}, \bibinfo {author} {\bibfnamefont
  {M.}~\bibnamefont {Hudl}}, \bibinfo {author} {\bibfnamefont {R.}~\bibnamefont
  {Neubert}}, \bibinfo {author} {\bibfnamefont {M.}~\bibnamefont {Thürk}},
  \bibinfo {author} {\bibfnamefont {S.}~\bibnamefont {Nietzsche}}, \bibinfo
  {author} {\bibfnamefont {W.}~\bibnamefont {Vodel}}, \bibinfo {author}
  {\bibfnamefont {P.}~\bibnamefont {Seidel}}, \ and\ \bibinfo {author}
  {\bibfnamefont {A.}~\bibnamefont {Tünnermann}},\ }\bibfield  {title}
  {\enquote {\bibinfo {title} {High mechanical {Q}-factor measurements on
  silicon bulk samples},}\ }\href@noop {} {\bibfield  {journal} {\bibinfo
  {journal} {J. Phys.: Conf. Ser.}\ }\textbf {\bibinfo {volume} {122}},\
  \bibinfo {pages} {012008} (\bibinfo {year} {2008})}\BibitemShut {NoStop}%
\bibitem [{\citenamefont {Gretarsson}\ and\ \citenamefont
  {Harry}(1999)}]{Gretarsson1999}%
  \BibitemOpen
  \bibfield  {author} {\bibinfo {author} {\bibfnamefont {A.~M.}\ \bibnamefont
  {Gretarsson}}\ and\ \bibinfo {author} {\bibfnamefont {G.~M.}\ \bibnamefont
  {Harry}},\ }\bibfield  {title} {\enquote {\bibinfo {title} {Dissipation of
  mechanical energy in fused silica fibers},}\ }\href@noop {} {\bibfield
  {journal} {\bibinfo  {journal} {Rev. Sci. Instrum.}\ }\textbf {\bibinfo
  {volume} {70}},\ \bibinfo {pages} {4081--4087} (\bibinfo {year}
  {1999})}\BibitemShut {NoStop}%
\bibitem [{\citenamefont {Ageev}\ \emph {et~al.}(2004)\citenamefont {Ageev},
  \citenamefont {Palmer}, \citenamefont {Felice}, \citenamefont {Penn},\ and\
  \citenamefont {Saulson}}]{Ageev2004}%
  \BibitemOpen
  \bibfield  {author} {\bibinfo {author} {\bibfnamefont {A.}~\bibnamefont
  {Ageev}}, \bibinfo {author} {\bibfnamefont {B.~C.}\ \bibnamefont {Palmer}},
  \bibinfo {author} {\bibfnamefont {A.~D.}\ \bibnamefont {Felice}}, \bibinfo
  {author} {\bibfnamefont {S.~D.}\ \bibnamefont {Penn}}, \ and\ \bibinfo
  {author} {\bibfnamefont {P.~R.}\ \bibnamefont {Saulson}},\ }\bibfield
  {title} {\enquote {\bibinfo {title} {Very high quality factor measured in
  annealed fused silica},}\ }\href@noop {} {\bibfield  {journal} {\bibinfo
  {journal} {Class. Quantum Grav.}\ }\textbf {\bibinfo {volume} {21}},\
  \bibinfo {pages} {3887--3892} (\bibinfo {year} {2004})}\BibitemShut {NoStop}%
\bibitem [{\citenamefont {Penn}\ \emph {et~al.}(2006)\citenamefont {Penn},
  \citenamefont {Ageev}, \citenamefont {Busby}, \citenamefont {Harry},
  \citenamefont {Gretarsson}, \citenamefont {Numata},\ and\ \citenamefont
  {Willems}}]{Penn2006}%
  \BibitemOpen
  \bibfield  {author} {\bibinfo {author} {\bibfnamefont {S.~D.}\ \bibnamefont
  {Penn}}, \bibinfo {author} {\bibfnamefont {A.}~\bibnamefont {Ageev}},
  \bibinfo {author} {\bibfnamefont {D.}~\bibnamefont {Busby}}, \bibinfo
  {author} {\bibfnamefont {G.~M.}\ \bibnamefont {Harry}}, \bibinfo {author}
  {\bibfnamefont {A.~M.}\ \bibnamefont {Gretarsson}}, \bibinfo {author}
  {\bibfnamefont {K.}~\bibnamefont {Numata}}, \ and\ \bibinfo {author}
  {\bibfnamefont {P.}~\bibnamefont {Willems}},\ }\bibfield  {title} {\enquote
  {\bibinfo {title} {Frequency and surface dependence of the mechanical loss in
  fused silica},}\ }\href@noop {} {\bibfield  {journal} {\bibinfo  {journal}
  {Phys. Lett. A}\ }\textbf {\bibinfo {volume} {352}},\ \bibinfo {pages} {3--6}
  (\bibinfo {year} {2006})}\BibitemShut {NoStop}%
\bibitem [{\citenamefont {Sun}\ \emph {et~al.}(2008)\citenamefont {Sun},
  \citenamefont {Wang}, \citenamefont {Li},\ and\ \citenamefont
  {Zhou}}]{Sun2008}%
  \BibitemOpen
  \bibfield  {author} {\bibinfo {author} {\bibfnamefont {Z.}~\bibnamefont
  {Sun}}, \bibinfo {author} {\bibfnamefont {J.}~\bibnamefont {Wang}}, \bibinfo
  {author} {\bibfnamefont {M.}~\bibnamefont {Li}}, \ and\ \bibinfo {author}
  {\bibfnamefont {Y.}~\bibnamefont {Zhou}},\ }\bibfield  {title} {\enquote
  {\bibinfo {title} {Mechanical properties and damage tolerance of
  {Y}$_2${S}i{O}$_5$},}\ }\href@noop {} {\bibfield  {journal} {\bibinfo
  {journal} {J. Eur. Ceram. Soc.}\ }\textbf {\bibinfo {volume} {28}},\ \bibinfo
  {pages} {2895--2901} (\bibinfo {year} {2008})}\BibitemShut {NoStop}%
\bibitem [{\citenamefont {Marion}\ and\ \citenamefont
  {Beach}(1990)}]{Marion1990}%
  \BibitemOpen
  \bibfield  {author} {\bibinfo {author} {\bibfnamefont {J.}~\bibnamefont
  {Marion}}\ and\ \bibinfo {author} {\bibfnamefont {R.}~\bibnamefont {Beach}},\
  }\bibfield  {title} {\enquote {\bibinfo {title} {Thermalphysical {P}roperties
  of {Y}$_2${S}i{O}$_5$ ({YOS})},}\ }\href@noop {} {\bibfield  {journal}
  {\bibinfo  {journal} {LRD}\ ,\ \bibinfo {pages} {90--038}} (\bibinfo {year}
  {1990})}\BibitemShut {NoStop}%
\bibitem [{Har()}]{Hartman2024}%
  \BibitemOpen
  \href@noop {} {\emph {\bibinfo {title} {To be published}}}\BibitemShut
  {NoStop}%
\bibitem [{\citenamefont {Galland}\ \emph {et~al.}(2020)\citenamefont
  {Galland}, \citenamefont {Lučić}, \citenamefont {Fang}, \citenamefont
  {Zhang}, \citenamefont {Le~Targat}, \citenamefont {Ferrier}, \citenamefont
  {Goldner}, \citenamefont {Seidelin},\ and\ \citenamefont
  {Le~Coq}}]{Galland2020}%
  \BibitemOpen
  \bibfield  {author} {\bibinfo {author} {\bibfnamefont {N.}~\bibnamefont
  {Galland}}, \bibinfo {author} {\bibfnamefont {N.}~\bibnamefont {Lučić}},
  \bibinfo {author} {\bibfnamefont {B.}~\bibnamefont {Fang}}, \bibinfo {author}
  {\bibfnamefont {S.}~\bibnamefont {Zhang}}, \bibinfo {author} {\bibfnamefont
  {R.}~\bibnamefont {Le~Targat}}, \bibinfo {author} {\bibfnamefont
  {A.}~\bibnamefont {Ferrier}}, \bibinfo {author} {\bibfnamefont
  {P.}~\bibnamefont {Goldner}}, \bibinfo {author} {\bibfnamefont
  {S.}~\bibnamefont {Seidelin}}, \ and\ \bibinfo {author} {\bibfnamefont
  {Y.}~\bibnamefont {Le~Coq}},\ }\bibfield  {title} {\enquote {\bibinfo {title}
  {Mechanical {T}unability of an {U}ltranarrow {S}pectral {F}eature of a
  {R}are-{E}arth-{D}oped {C}rystal via {U}niaxial {S}tress},}\ }\href@noop {}
  {\bibfield  {journal} {\bibinfo  {journal} {Phys. Rev. Applied}\ }\textbf
  {\bibinfo {volume} {13}},\ \bibinfo {pages} {044022} (\bibinfo {year}
  {2020})}\BibitemShut {NoStop}%
\bibitem [{\citenamefont {Mirzai}\ \emph {et~al.}(2021)\citenamefont {Mirzai},
  \citenamefont {Ahadi}, \citenamefont {Melin},\ and\ \citenamefont
  {Olsson}}]{Mirzai2021}%
  \BibitemOpen
  \bibfield  {author} {\bibinfo {author} {\bibfnamefont {A.}~\bibnamefont
  {Mirzai}}, \bibinfo {author} {\bibfnamefont {A.}~\bibnamefont {Ahadi}},
  \bibinfo {author} {\bibfnamefont {S.}~\bibnamefont {Melin}}, \ and\ \bibinfo
  {author} {\bibfnamefont {P.}~\bibnamefont {Olsson}},\ }\bibfield  {title}
  {\enquote {\bibinfo {title} {First-principle investigation of doping effects
  on mechanical and thermodynamic properties of {Y}$_2${S}i{O}$_5$},}\
  }\href@noop {} {\bibfield  {journal} {\bibinfo  {journal} {Mechanics of
  Materials}\ }\textbf {\bibinfo {volume} {154}},\ \bibinfo {pages} {103739}
  (\bibinfo {year} {2021})}\BibitemShut {NoStop}%
\bibitem [{\citenamefont {Cook}, \citenamefont {Rosenband},\ and\ \citenamefont
  {Leibrandt}(2015)}]{Cook2015}%
  \BibitemOpen
  \bibfield  {author} {\bibinfo {author} {\bibfnamefont {S.}~\bibnamefont
  {Cook}}, \bibinfo {author} {\bibfnamefont {T.}~\bibnamefont {Rosenband}}, \
  and\ \bibinfo {author} {\bibfnamefont {D.~R.}\ \bibnamefont {Leibrandt}},\
  }\bibfield  {title} {\enquote {\bibinfo {title} {{L}aser-{F}requency
  {S}tabilization {B}ased on {S}teady-{S}tate {S}pectral-{H}ole {B}urning in
  {E}u$^{3+}$:{Y}$_2${S}i{O}$_5$},}\ }\href@noop {} {\bibfield  {journal}
  {\bibinfo  {journal} {Phys. Rev. Lett.}\ }\textbf {\bibinfo {volume} {114}},\
  \bibinfo {pages} {253902} (\bibinfo {year} {2015})}\BibitemShut {NoStop}%
\bibitem [{\citenamefont {Lin}\ \emph {et~al.}(2024)\citenamefont {Lin},
  \citenamefont {Hartman}, \citenamefont {Pointard}, \citenamefont {Goldner},
  \citenamefont {Seidelin}, \citenamefont {Fang},\ and\ \citenamefont
  {Le~Coq}}]{Lin2024}%
  \BibitemOpen
  \bibfield  {author} {\bibinfo {author} {\bibfnamefont {X.}~\bibnamefont
  {Lin}}, \bibinfo {author} {\bibfnamefont {M.~T.}\ \bibnamefont {Hartman}},
  \bibinfo {author} {\bibfnamefont {B.}~\bibnamefont {Pointard}}, \bibinfo
  {author} {\bibfnamefont {P.}~\bibnamefont {Goldner}}, \bibinfo {author}
  {\bibfnamefont {S.}~\bibnamefont {Seidelin}}, \bibinfo {author}
  {\bibfnamefont {B.}~\bibnamefont {Fang}}, \ and\ \bibinfo {author}
  {\bibfnamefont {Y.}~\bibnamefont {Le~Coq}},\ }\bibfield  {title} {\enquote
  {\bibinfo {title} {Anomalous sub-kelvin thermal frequency shifts of ultra
  narrow-lindwidth solid state emitters},}\ }\href@noop {} {\bibfield
  {journal} {\bibinfo  {journal} {Phys. Rev. Lett.}\ ,\ \bibinfo {pages} {in
  press}} (\bibinfo {year} {2024})}\BibitemShut {NoStop}%
\bibitem [{\citenamefont {Yano}, \citenamefont {Mitsunaga},\ and\ \citenamefont
  {Uesugi}(1991)}]{Yano1991}%
  \BibitemOpen
  \bibfield  {author} {\bibinfo {author} {\bibfnamefont {R.}~\bibnamefont
  {Yano}}, \bibinfo {author} {\bibfnamefont {M.}~\bibnamefont {Mitsunaga}}, \
  and\ \bibinfo {author} {\bibfnamefont {N.}~\bibnamefont {Uesugi}},\
  }\bibfield  {title} {\enquote {\bibinfo {title} {Ultralong optical dephasing
  time in {E}r$^{3+}$:{Y}$_2${S}i{O}$_5$},}\ }\href@noop {} {\bibfield
  {journal} {\bibinfo  {journal} {Opt. Lett.}\ }\textbf {\bibinfo {volume}
  {16}},\ \bibinfo {pages} {1884} (\bibinfo {year} {1991})}\BibitemShut
  {NoStop}%
\bibitem [{\citenamefont {Yano}, \citenamefont {Mitsunaga},\ and\ \citenamefont
  {Uesugi}(1992)}]{Yano1992}%
  \BibitemOpen
  \bibfield  {author} {\bibinfo {author} {\bibfnamefont {R.}~\bibnamefont
  {Yano}}, \bibinfo {author} {\bibfnamefont {M.}~\bibnamefont {Mitsunaga}}, \
  and\ \bibinfo {author} {\bibfnamefont {N.}~\bibnamefont {Uesugi}},\
  }\bibfield  {title} {\enquote {\bibinfo {title} {Nonlinear laser spectroscopy
  of {E}r$^{3+}$:{Y}$_2${S}i{O}$_5$ and its application to time-domain optical
  memory},}\ }\href@noop {} {\bibfield  {journal} {\bibinfo  {journal} {J. Opt.
  Soc. Am. B}\ }\textbf {\bibinfo {volume} {9}},\ \bibinfo {pages} {992}
  (\bibinfo {year} {1992})}\BibitemShut {NoStop}%
\bibitem [{\citenamefont {Matei}\ \emph {et~al.}(2017)\citenamefont {Matei},
  \citenamefont {Legero}, \citenamefont {Häfner}, \citenamefont {Grebing},
  \citenamefont {Weyrich}, \citenamefont {Zhang}, \citenamefont {Sonderhouse},
  \citenamefont {Robinson}, \citenamefont {Ye}, \citenamefont {Riehle},\ and\
  \citenamefont {Sterr}}]{Matei2017}%
  \BibitemOpen
  \bibfield  {author} {\bibinfo {author} {\bibfnamefont {D.~G.}\ \bibnamefont
  {Matei}}, \bibinfo {author} {\bibfnamefont {T.}~\bibnamefont {Legero}},
  \bibinfo {author} {\bibfnamefont {S.}~\bibnamefont {Häfner}}, \bibinfo
  {author} {\bibfnamefont {C.}~\bibnamefont {Grebing}}, \bibinfo {author}
  {\bibfnamefont {R.}~\bibnamefont {Weyrich}}, \bibinfo {author} {\bibfnamefont
  {W.}~\bibnamefont {Zhang}}, \bibinfo {author} {\bibfnamefont
  {L.}~\bibnamefont {Sonderhouse}}, \bibinfo {author} {\bibfnamefont {J.~M.}\
  \bibnamefont {Robinson}}, \bibinfo {author} {\bibfnamefont {J.}~\bibnamefont
  {Ye}}, \bibinfo {author} {\bibfnamefont {F.}~\bibnamefont {Riehle}}, \ and\
  \bibinfo {author} {\bibfnamefont {U.}~\bibnamefont {Sterr}},\ }\bibfield
  {title} {\enquote {\bibinfo {title} {1.5 $\mu$m {L}asers with {S}ub-10 m{H}z
  {L}inewidth},}\ }\href@noop {} {\bibfield  {journal} {\bibinfo  {journal}
  {Phys. Rev. Lett.}\ }\textbf {\bibinfo {volume} {118}},\ \bibinfo {pages}
  {263202} (\bibinfo {year} {2017})}\BibitemShut {NoStop}%
\end{thebibliography}%

\end{document}